\documentclass[twocolumn,dvipsnames]{aastex631}
\usepackage{graphicx}
\usepackage{xcolor}
\usepackage{amsmath}

\submitjournal{ApJ}

\newcommand{\ngc}{NGC~1453}
\newcommand{\mdm}{\ensuremath{M_{15}}}
\newcommand{\kms}{\ensuremath{}{\rm \, km~s^{-1}}}

\newcommand{\mbh}{\ensuremath{M_\mathrm{BH}}}
\newcommand{\ml}{\ensuremath{M^*/L_{\rm F110W}}}

\newcommand{\Tmaj}{\ensuremath{T_\mathrm{maj}}}
\newcommand{\Tmin}{\ensuremath{T_\mathrm{min}}}

\shorttitle{Triaxial Orbit Modelling}
\shortauthors{Quenneville et al.}

\begin{document}

\title{Triaxial Orbit-based Dynamical Modeling of Galaxies with Supermassive Black Holes and an Application to Massive Elliptical Galaxy NGC~1453}

\correspondingauthor{Matthew E. Quenneville}
\email{mquenneville@berkeley.edu}

\author{Matthew E. Quenneville}
\affiliation{Department of Astronomy, University of California, Berkeley, CA 94720, USA}
\affiliation{Department of Physics, University of California, Berkeley, CA 94720, USA}

\author{Christopher M. Liepold}
\affiliation{Department of Astronomy, University of California, Berkeley, CA 94720, USA}
\affiliation{Department of Physics, University of California, Berkeley, CA 94720, USA}

\author{Chung-Pei Ma}
\affiliation{Department of Astronomy, University of California, Berkeley, CA 94720, USA}
\affiliation{Department of Physics, University of California, Berkeley, CA 94720, USA}

\begin{abstract}

Most stellar-dynamical determinations of the masses of nearby supermassive black holes (SMBHs) have been obtained with the orbit superposition technique under the assumption of axisymmetry. However, few galaxies -- in particular massive early-type galaxies -- obey exact axisymmetry. Here we present a revised orbit superposition code and a new approach for 
dynamically determining the intrinsic shapes and mass parameters of triaxial galaxies
based on spatially-resolved stellar kinematic data. 
The triaxial TriOS code described here corrects an error in the original \citet{vandenBoschetal2008} code that
gives rise to incorrect projections for most orbits in triaxial models and can significantly impact parameter search results. The revised code also contains significant improvements in orbit sampling, mass constraints, and run time. 
Furthermore, we introduce two new parameter searching strategies -- a new set of triaxial shape parameters and a novel grid-free sampling technique -- that together lead to a remarkable gain in efficiency in locating the best-fit model.
We apply the updated code and search method to NGC~1453, a fast-rotating massive elliptical galaxy. A full 6D parameter search finds $p=b/a=0.933^{+0.014}_{-0.015}$ and $q=c/a=0.779\pm0.012$ for the intrinsic axis ratios and $T=0.33\pm0.06$ for the triaxiality parameter. 
Despite the deviations from axisymmetry, the best-fit SMBH mass, stellar mass-to-light ratio, and dark matter enclosed mass for NGC~1453 are consistent with the axisymmetric results. More comparisons between axisymmetric and triaxial modeling are needed before drawing general conclusions.
        
\end{abstract}


\section{Introduction}
\label{sec:introduction}

Elliptical galaxies exhibit a wide range of isophotal shapes and surface brightness profiles.
There is an intrinsic uncertainty in inferring the 3D stellar luminosity density from the observed 2D isophotes on the sky.
When stellar kinematics from spectroscopic observations are combined with photometric information,
stronger constraints can be placed on the intrinsic 3D shapes of elliptical galaxies (e.g., \citealt{Binney1985,Franxetal1991}).
An idealized galaxy obeying exact axisymmetry would, by construction, have a regular surface brightness distribution without any isophotal twists and have perfectly aligned photometric and kinematic axes. Triaxial systems, on the other hand, can have isophotal twists, misaligned photometric and kinematic axes, and other spatially varying kinematic features absent in an axisymmetric system.
This consideration led \citet{Binney1985} to argue that triaxiality is common among elliptical galaxies.

Since then, a more detailed picture has emerged. Elliptical galaxies with lower stellar mass ($M_* \lesssim 10^{11.5} M_\odot$) tend to exhibit properties typical of axisymmetry \citep[e.g.,][]{emsellemetal2007, Weijmansetal2014, Cappellari2016,  Fosteretal2017}. Comparatively, elliptical galaxies with higher mass ($M_* \gtrsim 10^{11.5} M_\odot$) typically exhibit photometric twists, slow or no rotation, and misalignments between the photometric and kinematic axes, suggesting triaxial intrinsic shapes \citep[e.g.,][]{Vealeetal2017a, Vealeetal2017b, Vealeetal2018, Krajnovicetal2018, Eneetal2018, Goullaudetal2018, Eneetal2020}.
Thus, it is vital to understand the role of triaxiality in dynamical galaxy modelling, particularly in studying massive elliptical galaxies and their central black holes in the local universe.

The most massive SMBHs observed in the nearby universe lie in centers of some of the most massive nearby elliptical galaxies \citep{Maetal2014}. However, few triaxial SMBH mass (\mbh{}) measurements have been published thus far, perhaps because of the complexity in orbital structures, high-dimensional parameter space, and the associated computational cost required to model stellar orbits in triaxial potentials.  To date, all published \mbh\ measurements based on triaxial orbit modeling have been performed using the code initially presented in \citet{vandenBoschetal2008}. 
This code was first applied to determine the intrinsic shapes and \mbh\ of two fast-rotating elliptical galaxies M32 and NGC 3379 \citep{vandenBoschdeZeeuw2010}.
In this work, M32 was found to be near oblate axisymmetry with $\mbh= (2.4\pm1.0)\times10^6 M_\odot$, fully consistent with \mbh\ from earlier axisymmetric models~\citep[][]{vanderMareletal1998,Josephetal2001,Verolmeetal2002}.  NGC~3379, on the other hand, was found to be moderately triaxial, and the inferred $\mbh= (4\pm1)\times10^8M_\odot$ was double the value derived from axisymmetric models~\citep[][]{Gebhardtetal2000a,Shapiroetal2006}.
In a subsequent application to the S0 galaxy NGC~3998 \citep{Walshetal2012}, the best-fit model was found to be moderately triaxial although oblate axisymmetry was not ruled out. 

\citet{FeldmeierKrause2017} applied
the \citet{vandenBoschetal2008} code to the nuclear star cluster and SMBH at the Galactic center. The cluster shape was strongly triaxial, and the inferred \mbh\ was consistent within $1\,\sigma$ of 
the values inferred from the orbit of the S2 star \citep{gravitycollaboration2019, Doetal2019}.

More recently, \citet{denBroketal2021} used the \citet{vandenBoschetal2008} code to model PGC 046832. This galaxy exhibits dramatic twists, and the resulting models preferred strong variations in triaxiality. However, while axisymmetric models suggested a central black hole mass of $6\times10^9M_\odot$, the triaxial models prefer models with no central black hole. Instead they report an upper bound on the central black hole mass of $2\times10^9M_\odot$. This differs significantly from the value determined from axisymmetric models.

In addition to these published triaxial \mbh\ values, the  \citet{vandenBoschetal2008} code has been used to determine several \mbh\ in the nearly axisymmetric limit \citep{Sethetal2014,Walshetal2015,Walshetal2016,Walshetal2017,Ahnetal2018}. It has also been used to estimate the intrinsic triaxiality of galaxies under the assumption of a fixed \mbh\ \citep[e.g.,][]{vandenBoschetal2008,Leungetal2018,Zhuetal2018a,Zhuetal2018b,Pocietal2019,Yangetal2020,Jinetal2020}.

We have been revamping the \citet{vandenBoschetal2008} code for a systematic study of the SMBHs and other mass components in the $\sim 100$ most massive local early-type galaxies in the MASSIVE survey \citep{Maetal2014}.
As a first step, we introduced a version of the code capable of achieving the exact axisymmetric limit \citep{Liepoldetal2020, Quennevilleetal2021}.  The original \citet{vandenBoschetal2008} code was (intentionally) not built to respect axisymmetry, but it had been used to perform (nearly) axisymmetric orbit modeling, leading to unexplained inconsistencies when the resulting \mbh\ values were compared to those from axisymmetric orbit codes (e.g., \citealt{Ahnetal2018}). Our axisymmetrized version of the code has bridged this gap and now enables dynamical modeling of galaxies using stellar orbits that properly obey axisymmetry. We applied our axisymmetrized code to NGC~1453, a fast-rotating elliptical galaxy in the MASSIVE survey, and obtained a significant detection of its SMBH with $\mbh=(2.9\pm0.4)\times10^9~M_\odot$ \citep{Liepoldetal2020}.  Models without black holes were excluded at the $8.7\sigma$ level. 

For clarity, we refer to the original code (which was unnamed) by the citation \citet{vandenBoschetal2008}, and refer to our versions as the TriOS (Triaxial Orbit Superposition) code.

In this paper, we move beyond the axisymmetric limit of \citet{Quennevilleetal2021}, and present a triaxial version of the TriOS code and a first application of this code.  This triaxial TriOS code differs in a number of major ways from the original \citet{vandenBoschetal2008} code. We have implemented these changes to correct a number of bugs and issues that we uncovered during extensive tests of the original code for triaxial potentials.
As a start, we correct a major error in the orbit construction part of the code that incorrectly flips some velocity components for the tube orbits. Our tests indicate that for most viewing angles, correcting this mistake has a significant impact on the resulting orbital kinematics and galaxy model parameter recovery within the code.
Other major changes include (i)
modifying the acceleration table used for orbit integration to gain
a significant speedup in runtime, (ii) resolving issues with insufficient orbit sampling that can result in spurious shape preferences, and (iii) using a more uniform mass binning scheme to eliminate frequent problems in satisfying mass constraints.  Details of these changes are described in Section~\ref{sec:modifications}.

In addition to these code changes, 
we introduce a new set of shape parameters in this paper (Section~\ref{sec:new_params}) that are chosen to improve the efficiency of parameter searches in triaxial galaxy shapes and orientations.  These parameters strike a balance between sampling in galaxy intrinsic shape and galaxy orientation, and result in fewer unrealistically flat galaxy shapes.
To place these new parameters in context, we provide a summary (Section~\ref{sec:triaxiality})
of the parameters used in previous work to describe a triaxial galaxy's intrinsic and observed axis ratios,
the relations of viewing angles and sky projections,
and how an observed surface brightness is deprojected to obtain a 3D intrinsic shape within the TriOS code. 

We apply our triaxial TriOS code to NGC~1453 in the final part of the paper (Section~\ref{sec:dyn_models}).
Since triaxial modeling typically involves at least five parameters (three for shapes and at least two for mass parameters), we introduce an efficient new search strategy for sampling this multi-dimensional parameter space. This new strategy does not rely on direct grid searches used in previous orbit modeling studies. Instead, we apply nested Latin hypercube sampling to a 6D parameter space and are able to converge to a best-fit model for NGC~1453 with an order-of-magnitude fewer sample points. The resulting best-fit triaxial model is compared to the best-fit axisymmetric model from \citet{Liepoldetal2020}. 

\section{Modeling a Triaxial Galaxy}
\label{sec:triaxiality}

In this section we summarize the information relevant for modeling a triaxial galaxy, e.g., coordinate systems, intrinsic and apparent shape parameters, viewing angles, and sky projections.

\subsection{Intrinsic Shapes and Axis Ratios}
\label{sec:shapes}

To describe the 3D structure of a galaxy, we use a Cartesian coordinate system centered at the galaxy's nucleus, in which the $x$, $y$, and $z$ axes are directed along the intrinsic major, intermediate, and minor axes of the galaxy, respectively.  The $z$-axis is therefore the symmetry axis of an oblate axisymmetric galaxy, and the $x$-axis is the symmetry axis of a prolate axisymmetric galaxy. 

It is convenient to use a different coordinate system to describe properties projected on the sky.  We follow the standard practice and take the $x'$  and $y'$ axes of this coordinate system to be along the major and minor axes of the projected surface brightness distribution of a galaxy. The $z'$ axis is along the line-of-sight.

We use $a$, $b$, $c$ to denote the lengths of the three principal axes of a triaxial ellipsoidal isodensity surface, assuming $c \le b \le a$.  We use $a'$ and $b'$ to denote the lengths of the (observed) major and minor axes of the projected ellipse on the sky. Four useful axis ratios are
\begin{equation}
\label{eq:axis_ratios}
    p=\frac{b}{a}\,, \quad  q=\frac{c}{a}\,, \quad  u=\frac{a'}{a}\,, \quad
    q'=\frac{b'}{a'}\,,
\end{equation}
where $p$ is the intrinsic intermediate-to-major axis ratio, $q$ is the intrinsic minor-to-major axis ratio, 
$u$ represents a compression factor between the intrinsic major axis and the apparent major axis on the sky due to projection, and $q'$ is the flattening of the projected shape.
These quantities obey the inequalities
\begin{eqnarray}
    &&   0 \le c \le b' \le b \le a' \le a \,, \nonumber \\
    & {\rm or}\quad  &  0 \le q \le uq' \le p \le u \le 1  \,.
\label{eq:inequality}
\end{eqnarray}
The upper and lower limits of $u$ correspond to the intrinsic major axis lying in the plane of the sky ($u=1$ or $a'=a$) and the intrinsic intermediate axis lying in the plane of the sky ($u=p$ or $a'=b$), respectively.

The commonly used triaxiality parameter is
\begin{equation}
    T=\frac{a^2-b^2}{a^2-c^2} = \frac{1-p^2}{1-q^2}\,,
    \label{eq:T}
\end{equation} 
which ranges between 0 for an oblate axisymmetric shape ($a=b$),
and 1 for a prolate axisymmetric shape ($b=c$), with values between 0 and 1 indicating a triaxial shape.

\subsection{Viewing Angles and Sky Projections}
\label{sec:angles}

A line of sight between an observer and a galaxy is specified by two viewing angles $(\theta, \phi)$, where $\theta$ and $\phi$ are the usual polar angles in the galaxy's intrinsic $(x, y, z)$ coordinate system. Thus, $\theta=0^\circ$ is for a line of sight along the intrinsic minor axis (i.e., a face-on view down the $z$-axis), and $\theta=90^\circ$ is for lines of sight in the $x-y$ plane (i.e., an edge-on view with the intrinsic minor axis in the sky plane). Similarly,  $\phi=0^\circ$ is for lines of sight in the $x-z$ plane (i.e., the intrinsic intermediate axis is in the sky plane), and $\phi=90^\circ$ is for lines of sight in the $y-z$ plane (i.e., the intrinsic major axis is in the sky plane).

Given a triaxial 3D density stratified on similar concentric ellipsoids,
the viewing angle $\theta$ and $\phi$ are sufficient to project the 3D shape and determine the 2D projected coordinate system $(x',y')$. 
To de-project an observed 2D shape on the sky, however, a third angle, $\psi$, is needed to completely specify the intrinsic coordinate system.
This third angle $\psi$ specifies the remaining degree of freedom once $\theta$ and $\phi$ are fixed -- a rotation of the galaxy around the line of sight.  More precisely, $\psi$ is defined as the angle between the $y'$ axis, and the line defined by the intersection of the $x'-y'$ and $x-y$ planes. When $\psi=0^\circ$, the $x-y$ plane and $x'-y'$ plane intersect along the $y'$ axis; when $\psi=90^\circ$, the $x-y$ plane and $x'-y'$ plane intersect along the $x'$ axis.  

Together, the three angles $(\theta, \phi, \psi)$ uniquely specify the orientation of the intrinsic axes with respect to the projected axes.  If the 3D density is stratified on similar concentric ellipsoidal surfaces, the axis ratios $(p, q, u)$ of Equation~(\ref{eq:axis_ratios}) can be uniquely determined from the projected surface brightness and $(\theta, \phi, \psi)$ using the equations from Appendix A of \citet{deZeeuwFranx1989}.

\subsection{Deprojecting Observed Surface Brightness}
\label{sec:deprojection}

Within the TriOS code, the 3D stellar density distribution is described by a sum of multiple Gaussian components of varying widths and axis ratios using the Multi-Gaussian Expansion (MGE) scheme \citep{Cappellari2002}. To determine these components, one first fits 
a 2D MGE to the observed surface brightness of the galaxy. Each MGE component is allowed to have its own projected flattening $q'$ to account for radially varying ellipticity in the observed isophotes. In addition, each MGE component can have a different position angle (PA) to accommodate any observed isophotal twists.

In general, the deprojection of a 2D surface brightness distribution to give a 3D triaxial luminosity density is not unique. MGE is a parametric method of choosing one particular 3D density for a given 2D surface brightness and set of intrinsic axes. Non-parametric deprojection methods have also recently been developed for triaxial galaxies in \citet{deNicolaetal2020}, but the TriOS code is not yet capable of using these deprojections.

For a set of $(\theta,\phi,\psi)$ that specifies the alignment of the galaxy's intrinsic principle axes $(x, y, z)$, one can determine the deprojection of each MGE component that shares these principle axes (if a valid deprojection exists). This deprojection is unique due to the assumption that each 2D gaussian corresponds with a 3D gaussian density with similar concentric ellipsoidal surfaces of constant density.
The axis ratios $p$ and $q$ of each deprojected MGE component can have their own values. 
The triaxiality parameter $T$, on the other hand, has the convenient property that it is identical for all MGE components when the components share the same PA (i.e., no isophotal twists).\footnote{This is valid as long as the line-of-sight does not lie in a principal plane of the triaxial shape.  If it does, then all aligned 3D ellipsoids will have parallel or perpendicular PAs when viewed in projection, and differences in $T$ cannot be inferred from differences in projected PA. We do not consider any models with lines-of-sight lying directly in the principal planes.}

\section{New Parameters for Triaxial Space Sampling}
\label{sec:new_params}

\subsection{Prior practice}

As discussed in Section~\ref{sec:triaxiality}, either $(p, q, u)$ or $(\theta, \phi, \psi)$ can be used to specify the shape of a triaxial galaxy and its sky projections.  One can in principle search in either space when running orbit models to determine a galaxy's intrinsic shape and mass parameters.  In practice, however, prior triaxial orbit modeling studies favored $(p, q, u)$ over the angles. In these studies, the orbit models were typically run for a grid of regularly spaced values of $(p, q, u)$ \citep[e.g.,][]{vandenBoschvandeVen2009, vandenBoschdeZeeuw2010, Walshetal2012, Jinetal2019}. In a few other triaxial studies, 
$u$ was fixed to some value close to 1 while the parameter search was conducted over $p$ and $q$ in a regular 2D grid \citep[e.g.,][]{Zhuetal2018a, Zhuetal2018b, Pocietal2019}.  Since $u \sim 1$ corresponds to the intrinsic major axis lying close to the sky plane, these studies did not search over all allowed viewing angles.

The argument used by \citet{vandenBoschvandeVen2009} for favoring conducting parameter searches in $(p, q, u)$ rather than $(\theta, \phi, \psi)$ is that a change in the angles can result in either a very small or very large change in axis ratios, depending on the angles being explored. We note, however, that the converse is also true: a change in the axis ratios can result in either a very small or very large change in the principal axes' alignment, depending on the values of these ratios. Two models with similar axis ratios, but viewed along very different lines of sight, can result in very different observables. An optimal sampling should consider both the intrinsic shape and the alignment of the line of sight.

\subsection{Properties of new parameters}
\label{sec:new_params_def}

Here we propose a new set of variables to parameterize a galaxy's intrinsic triaxial shape and its sky projections. The advantages of conducting parameter searches in these variables over either $(p, q, u)$ or $(\theta, \phi, \psi)$ during triaxial orbit modeling will be discussed in Section~\ref{sec:new_params_adv}.
 
For the first shape parameter, we choose the triaxiality parameter $T$ (Equation~\ref{eq:T}). 
We define the next two parameters with forms analogous to $T$:
\begin{equation}
\label{eq:shape_params}
\begin{split}
  T &=\frac{a^2-b^2}{a^2-c^2} = \frac{1-p^2}{1-q^2}\,,\\
    \Tmaj& \equiv \frac{a^2-a'^2}{a^2-b^2}
        =\frac{1-u^2}{1-p^2} \,, \\
    \Tmin& \equiv \frac{b'^2-c^2}{b^2-c^2}
       =\frac{(uq')^2-q^2}{p^2-q^2} \,, 
\end{split}
\end{equation}
where $\Tmaj$ parameterizes the length of the projected major axis, $a'$, relative to its allowed limits $a$ and $b$, and 
$\Tmin$ parameterizes the length of the projected minor axis, $b'$, relative to its allowed limits $b$ and $c$.  It then follows from the inequalities in Equation~(\ref{eq:inequality}) that $(T, \Tmaj, \Tmin)$ form a unit cube, i.e.,
\begin{equation}
\begin{split}
 0 & \le T \le 1\,, \\
 0 & \le \Tmaj \le 1\,, \\
 0 & \le \Tmin \le 1\,. 
\end{split}
\label{eq:Trange}
\end{equation}
The limiting cases represented by each face of the unit cube has the following physical significance: (i) $T=0$ and 1 correspond to oblate axisymmetric ($a=b$ or $p=1$) and prolate axisymmetric ($b=c$ or $p=q$) shapes, respectively; (ii) $\Tmaj=0$ and 1 correspond to 
the intrinsic major axis lying in the sky plane ($a'=a$ or $u=1$) and the intrinsic intermediate axis lying in the sky plane ($a'=b$ or $u=p$), respectively; (iii) $\Tmin=0$ and 1 correspond to the intrinsic minor axis lying in the sky plane ($b'=c$ or $uq'=q$) and the intrinsic intermediate axis lying in sky ($b'=b$ or $uq'=p$), respectively.
While both the $\Tmaj=1$ and $\Tmin=1$ planes correspond to the intrinsic intermediate axis in the sky plane, they represent two complementary ranges of viewing angles such that $b$ is equal to the projected {\it major} axis $a'$ for $\Tmaj=1$, whereas $b$ is equal to the projected {\it minor} axis $b'$ for $\Tmin=1$.

Equation~(\ref{eq:shape_params}), along with the requirement that $q>0$, yields the inequality
 \begin{equation}
    \frac{(1-T)\Tmin}{1-T\Tmaj} < q'^2 \,,
 \end{equation}
implying that for an observed axis ratio $q'$ on the sky, only the $(T, \Tmaj, \Tmin)$ region satisfying the inequality has valid deprojections.  When the projected shape is flattened ($q'<1$), some models within the unit cube will result in negative (and thus invalid) values of the squared minor axis length, $c^2$.  This volume surrounds the line $(T,\Tmaj,\Tmin)=(T,1,1)$, which does not have a valid deprojection for any flattened projected shape.

\subsection{Relating $(T, \Tmaj, \Tmin)$ to old parameters}

While Equations~(\ref{eq:shape_params}) relate our new parameters to $(p, q, u)$, it is often useful to do the inverse and convert a given set of $(T, \Tmaj, \Tmin)$ to $(p, q, u)$. To do so, we use these sequential expressions
\begin{equation} 
\label{eq:projection}
\begin{split}
    1-q^2&=\frac{1-q'^2}{1-(1-T)\Tmin-q'^2T\Tmaj} \,, \\
    1-p^2&=T(1-q^2) \,, \\
    1-u^2&=\Tmaj(1-p^2) \,. \\
\end{split}
\end{equation}
For a given set of $(T, \Tmaj, \Tmin)$, these equations define the deprojection from an observed MGE component with flattening, $q'$, to its 3D shape parameters, $(p, q, u)$.

Similarly, it is useful to convert $(T, \Tmaj, \Tmin)$ to the angles $(\theta,\phi,\psi)$:
\begin{equation}
\label{eq:angles_Ts}
    \begin{split}
    \cos^2{\theta} &= \Tmin (1-T\, \Tmaj) \,, \\
    \tan^2{\phi} &=\frac{1 - \Tmaj}{\Tmaj}
    \frac{1-\Tmin}{1 - \Tmin(1-T)}\,, \\
    \tan^2{\psi} &= \frac{\left[1-\Tmin(1-T)\right](1-T\,\Tmaj)(1- \Tmin)}
     {T^2(1-\Tmaj) \Tmaj\, \Tmin}  \,.
    \end{split}
\end{equation}
We choose to use the branch where $0^\circ \le \theta \le 90^\circ$, $0^\circ \le \phi \le 90^\circ$, and $90^\circ \le \psi \le 180^\circ$, though other equivalent branches exist as well.\footnote{For instance, if one prefers $0^\circ \le \psi \le 90^\circ$ and $0^\circ \le \theta \le 90^\circ$, then $\phi$ obeys  $-90^\circ \le \phi \le 0^\circ$. }
The inverse expressions relating $(T, \Tmaj, \Tmin)$ and $(\theta,\phi,\psi)$ are given in Appendix~\ref{sec:ap_shapes}.

Equations~(\ref{eq:projection}) and (\ref{eq:angles_Ts}), as well as Equations~(\ref{eq:angles_Ts_seq}) and (\ref{eq:angles_Ts_inv}), follow directly from the definitions in Equation~\eqref{eq:shape_params}, and the general expressions for the deprojection of a triaxial density that is stratified on similar, concentric ellipsoids \citep[e.g.,][discussed further in appendix \ref{sec:ap_shapes}]{deZeeuwFranx1989}. Furthermore, since Equation~(\ref{eq:angles_Ts}) and its inverse Equation~(\ref{eq:angles_Ts_inv}) make no reference to the observed flattening, the same set of values $(T,\Tmaj,\Tmin)$ can be used for a density that is composed of multiple such components with different flattening values. Thus, $(T,\Tmaj,\Tmin)$ and $(\theta,\phi,\psi)$ are simply different parameterizations of the same space. Equations (\ref{eq:projection}) and (\ref{eq:angles_Ts}), along with the equations listed in Appendix~\ref{sec:ap_shapes}, make no reference to the MGE formalism and are 
applicable to any triaxial system that meets these conditions. The existence and uniqueness of a valid deprojection are not affected by the choice of shape space parameterization outside the principal planes.

\begin{figure}
  \centering
  \includegraphics[width=\linewidth]{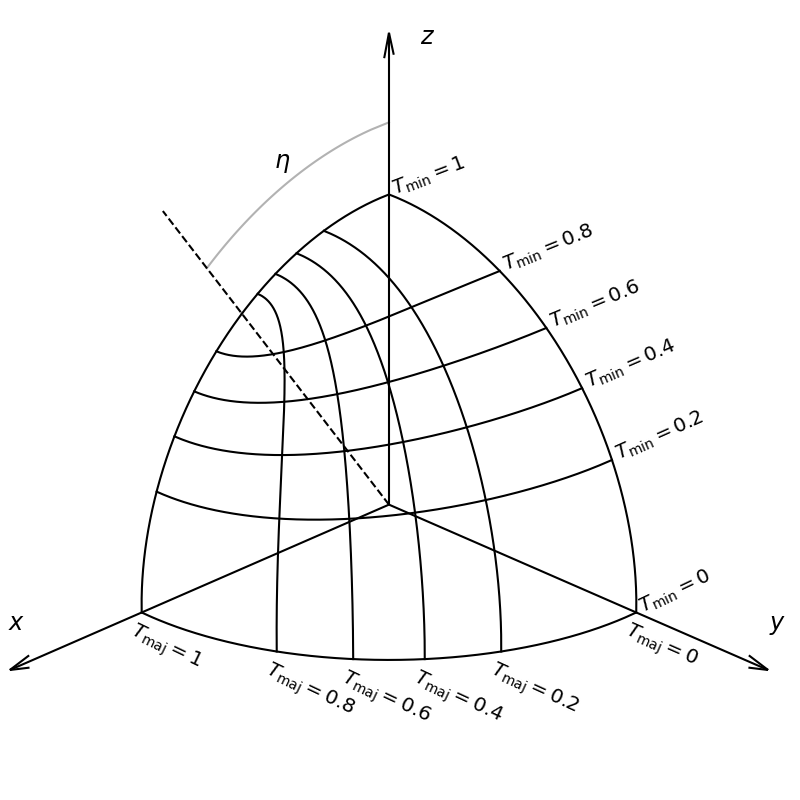}
  \hfill

  \caption{
  Isocontours of the new shape parameters, $\Tmaj$ and $\Tmin$, in a galaxy's coordinate system, where the $x$, $y$, and $z$ axes are chosen to be the intrinsic major, intermediate, and minor axes, respectively.  The triaxiality parameter, $T$, is assumed to be $0.35$ here. 
  The parameters \Tmaj\ and \Tmin\ are seen to change relatively uniformly with the line-of-sight direction, resulting in fewer unrealistically flattened models near non-deprojectable regions (see text).
  \label{figure:coordinates}}
\end{figure}

To illustrate the properties of \Tmaj\ and \Tmin, we plot a set of lines of constant \Tmaj\ and \Tmin\ in a galaxy's intrinsic coordinate system $x, y$, and $z$ in Figure~\ref{figure:coordinates}.  The corner points $(\Tmaj, \Tmin)=(1, 0)$, $(0, 0)$, and $(0, 1)$ correspond to viewing angles along the short, intermediate, and long axes, respectively. The point $(\Tmaj, \Tmin)=(1, 1)$ represents a line of sight lying along the line $\theta=\eta=\tan^{-1}(\sqrt{T/(1-T)})$ in the $x-z$ plane, which only results in a valid model for round projected shapes. For flattened shapes, there are no valid deprojections for lines of sight within a solid angle surrounding this direction. This non-deprojectable region increases in size, as the projected shape becomes flatter.

\subsection{Advantages of $T, \Tmaj$ and $\Tmin$}
\label{sec:new_params_adv}

The parameters $T$, \Tmaj, and \Tmin\ have a number of desirable properties. First, as Figure~\ref{figure:coordinates} illustrates, \Tmaj\ and \Tmin\ change relatively uniformly 
with the line-of-sight direction.
This is in contrast to the axis ratio space, $(p,q, u)$, in which tiny changes can result in large differences in the angles. 
For example, models with $p=0.99$ and a fixed $q$ would undergo a $90^\circ$ rotation in $\phi$ when $u$ is varied from 0.99 to 1.

Similarly, the galaxy shape varies much more uniformly with $(T, \Tmaj, \Tmin)$ than with $(\theta, \phi,\psi)$. Again, tiny changes in the latter can result in large differences in galaxy shape.  
For example, 
when an observed surface brightness (without isophotal twists) is deprojected into a 3D ellipsoidal shape with principle axes defined by $(\theta, \phi, \psi)=(89^\circ, 45^\circ, 90^\circ)$, Equation~(A2) shows that the resulting 3D shape has $T=0$, i.e., it is oblate axisymmetric.  As $\psi$ is increased from $90^\circ$ by only $\sim 1^\circ$, however, the deprojected shape varies drastically, with oblate axisymmetry at $\psi=90^\circ$ to prolate axisymmetry at $\psi\sim91^\circ$, with the full range of triaxialities lying in between. From Equation~(A2) with $\phi=45^\circ$, we find prolate axisymmetry ($T=1$) to occur when $\psi-90^\circ=(90^\circ/\pi)\arctan{\left(2\cos{\theta}/\sin^2{\theta}\right)}$ on our chosen branch. As $\theta$ approaches $90^\circ$, the value of $\psi$ that gives prolate axisymmetry approaches $90^\circ$. For $\theta=89^\circ$ (and $\phi=45^\circ$), prolate axisymmetry occurs at $\psi=90.99985^\circ$.

The behavior in the example above arises from coordinate singularities in the $(\theta,\phi,\psi)$ space. When the line-of-sight is chosen to lie in a principal plane (i.e., $\cos{(\theta)}=0,1$ or $\sin{(2\phi)}=0$), it is impossible for continuous photometric twists to arise in projection as triaxiality is varied. One consequence of this is that the only valid values of $\psi$ are $0^\circ$ or $90^\circ$, meaning it is no longer an independent parameter. Thus, $(\theta,\phi,\psi)$ are insufficient to fully specify the 3D projection. The parameters $(T,\Tmaj,\Tmin)$, on the other hand, have no such singularity. In the above example, the proximity of the chosen value of $\theta$ to $90^\circ$ causes the rapid shift in shape with $\psi$.

Another desirable property of \Tmaj\ and \Tmin\ is that, similar to $T$ (see Section~\ref{sec:deprojection}), they do not vary among MGE components with different axis ratios, so long as there are no isophotal twists. 

This invariant property can be explained by identifying 
\Tmaj\ and \Tmin\ as the shifted and rescaled versions of the conical coordinates,
$\mu_\mathrm{pro}$ and $\nu_\mathrm{pro}$
within the galaxy's intrinsic coordinate system \citep{Franx1988}, where $\mu_\mathrm{pro} = a'^2$ and $\nu_\mathrm{pro} = b'^2$.
Since the coordinate surfaces of $\mu_\mathrm{pro}$ and $\nu_\mathrm{pro}$ are the same for all MGE components, the shifted and scaled quantities $\Tmaj$ and $\Tmin$ do not vary between components.

The advantages of $T$, \Tmaj, and \Tmin\ are especially clear for systems not far from axisymmetry. Towards oblate axisymmetry ($T\approx 0$), we have $\Tmin \approx \cos^2{\theta}$ and $\Tmaj \approx \cos^2{\phi}$. Thus, a uniform sampling in $\sqrt{\Tmin}$ and $\sqrt{\Tmaj}$ will result in a nearly uniform sampling in the cosines of the inclination and the azimuthal angle. The same behavior holds towards prolate axisymmetry ($T\approx 1$) since the roles of $\Tmaj$ and $\Tmin$ are simply switched if the $x$ and $z$ axis labels are interchanged.
Thus, for nearly axisymmetric galaxies, a uniform sampling in $(T, \sqrt{\Tmaj}, \sqrt{\Tmin})$ results in fewer unrealistically flattened models.

\begin{table*}[]
\centering
\begin{tabular}{cccc}
\hline  \hline
Position     & Long-axis tube               & Short-axis tube               & Intermediate-axis tube             \\ \hline
$(x,y,z)$    & $(v_x,v_y,v_z)$          & $(v_x,v_y,v_z)$          & $(v_x,v_y,v_z)$                \\
$(-x,y,z)$   & $(-v_x,v_y,v_z)$         & $(v_x,-v_y,{\bf -v_z})$  & $(v_x,{\bf -v_y},-v_z)$        \\
$(x,-y,z)$   & $({\bf -v_x},v_y,-v_z)$  & $(-v_x,v_y,{\bf -v_z})$  & $(v_x,-v_y,v_z)$               \\
$(x,y,-z)$   & $({\bf -v_x},-v_y,v_z)$  & $(v_x,v_y,-v_z)$         & $(-v_x,{\bf -v_y},v_z)$        \\
$(-x,-y,z)$  & $({\bf v_x},v_y,-v_z)$   & $(-v_x,-v_y,v_z)$        & $(v_x,{\bf v_y},-v_z)$         \\
$(-x,y,-z)$  & $({\bf v_x},-v_y,v_z)$   & $(v_x,-v_y,{\bf v_z})$   & $(-v_x,v_y,-v_z)$              \\
$(x,-y,-z)$  & $(v_x,-v_y,-v_z)$        & $(-v_x,v_y,{\bf v_z})$   & $(-v_x,{\bf v_y},v_z)$         \\
$(-x,-y,-z)$ & $(-v_x,-v_y,-v_z)$       & $(-v_x,-v_y,-v_z)$       & $(-v_x,-v_y,-v_z)$             \\ \hline
\end{tabular}
\caption{Corrected mirroring scheme of the three types of tube orbits in our TriOS code.  Boldfaced velocity components have the opposite signs from the original scheme in Table~2 of \citet{vandenBoschetal2008}.
These components were flipped incorrectly in the original code.
\label{tab:flips}}
\end{table*}

\begin{figure*}
  \centering
  \includegraphics[width=\linewidth]{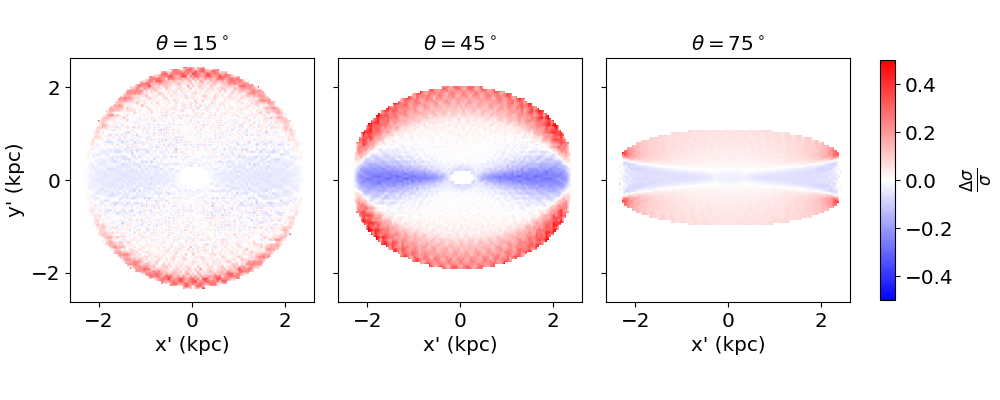}
  \hfill

  \caption{Illustration of the impact of the incorrect mirroring scheme in the \citet{vandenBoschetal2008} code.  We plot the fractional error between the incorrect and corrected schemes (see Table~1) in the kinematic map of the line-of-sight velocity dispersion, $\sigma$, for a single orbit. The orbit is chosen from the $x-z$ start space of a triaxial model with $T=6\times10^{-6}$ for NGC~1453, but it is representative of typical short-axis tubes in a triaxial potential.  Each panel represents a different viewing inclination angle $\theta$.  The fractional error is largest near $\theta=45^\circ$, reaching beyond 50\% for some parts of the orbit. 
  \label{figure:flips}}
\end{figure*}

\section{Code Corrections and Improvements}
\label{sec:modifications}

In this section, we describe the key corrections, improvements, and speedups made to the \citet{vandenBoschetal2008} code. See Section~4 of \citet{Quennevilleetal2021} for other general changes that we had implemented (regardless of axisymmetry).

\subsection{Correct orbital mirroring mistakes}
\label{sec:flips}
The TriOS code is written for a static triaxial potential that is symmetric under reflection along each of the three principal axes of a triaxial system. Under this assumption, any orbital property only needs to be calculated in one octant of the orbit space; it can then be ``mirrored'' into the other seven octants by symmetry. 

Taking advantage of this symmetry, the code initializes orbits in only one octant ($x,y,z >0$) and integrates only these orbits. Seven additional copies of each orbit are then created by simply mirroring along the three axes. The recipe for how to flip the signs of the velocity components is given in Table~2 of \citet{vandenBoschetal2008}. The exact procedure depends on whether the orbit is a short-axis tube, long-axis tube, or box.  These orbits are classified as follow: throughout its trajectory, an orbit is labelled a box orbit if all three components of its angular momentum $(L_x,L_y,L_z)$ change sign, and a tube orbit if exactly one component of angular momentum maintains its sign.  The tube orbits are further classified according to the angular momentum component that maintains its sign, i.e., a long-axis (i.e. $x$-axis) tube maintains the sign of its $L_x$, an intermediate-axis ($y$-axis) tube maintains the sign of $L_y$, and a short-axis ($z$-axis) tube maintains the sign of $L_z$. Orbits that don't fall into the tube or box orbit classifications are flipped in the same way as box orbits.

We discovered that the tube orbits are incorrectly flipped for four of the eight octants in Table~2 of \citet{vandenBoschetal2008}. We indicate the incorrect components in boldface and give the corrected expressions in Table~\ref{tab:flips}. The mistakes are such that the mirrored positions and velocities are inconsistent with one another, and the two do not combine to give a valid trajectory. 
A consequence of these mistakes is that the magnitude of each component of $\vec{L}$ is not always preserved by the mirroring, as it should be, and the resulting $|\vec{L}|$ is also not preserved. 
For instance, for the short-axis tube flip, the original recipe would change the amplitudes of $L_x$ and $L_y$ for 4 of the 8 copies, and the resulting total $L$ would not be preserved. Similarly, $L_y$ and $L_z$ are incorrect for 4 copies of the long-axis tubes, and $L_x$ and $L_z$ are incorrect for 4 copies of the intermediate-axis tubes. 

To illustrate the impact of the incorrect orbital flips, we plot the error in the line-of-sight velocity dispersion, $\sigma$, for a single short-axis tube orbit for three different viewing angles in Figure~\ref{figure:flips}. 
We first integrate the trajectory of this orbit within the potential and then compute the 7 mirrored copies using the original and corrected flips in Table~\ref{tab:flips}.  
The fractional difference in the projected $\sigma$ between the two schemes is then plotted 
for three different values of viewing inclination angle $\theta$. The errors vary across the plane of the sky and exceed 50\% for $\theta=45^\circ$.
For this orbit, the incorrect flip scheme tends to under-predict $\sigma$ along the galaxy's projected major axis and over-predict $\sigma$ near the edges.
The orbit shown in Figure~\ref{figure:flips} is typical of short-axis tubes in triaxial potentials. Long-axis tube orbits exhibit similar error patterns when the appropriate axis labels are switched.
While the pattern of velocity dispersion error is different for each orbit, systematic errors with magnitudes of $10\%-100\%$ are typical, with peak errors of over 1600\% in some cases for orbital inclinations near $45^\circ$.

To assess further the impact of the incorrect flips, we perform full orbit modeling for a grid of triaxial models for NGC~1453 using the original and then the corrected scheme. Overall, when the correct flips are used, we find that $\chi^2$ is lowered by a wide range of values depending on the triaxiality and viewing angles. For instance, the value of $\chi^2$ can decrease by more than 100 for strongly triaxial models, while it can change by less than 5 or even increase slightly for other models.  The overall $\chi^2$ landscape is therefore significantly altered by our corrections.

Due to the symmetry of the tube orbits, the errors in the orbital flips can cancel out when the galaxy is viewed along a principal axis.  Nearly axisymmetric models that are viewed edge-on or face-on will be similarly unaffected. Outside of these special cases, the orbital kinematics have significant errors.   The incorrect flips were not used in our axisymmetric modelling of NGC~1453 \citep{Liepoldetal2020, Quennevilleetal2021} since we used an axisymmetrization procedure in place of the flips in the TriOS code.

The discussion above is relevant only for tube orbits.  For box orbits, we find the flips given in Table~2 of \citet{vandenBoschetal2008} to be correct. However, in addition to this set of 8 mirrored orbits, we choose to include 8 more orbits for each point in the stationary start space (defined in section~\ref{sec:1453_sampling}) that correspond to enforcing time reversal symmetry for the box orbits. This addition ensures that box orbits have the expected even parity in their line-of-sight velocity distributions (LOSVDs).  In the cases that we have examined, these orbits already have small enough odd LOSVD components that this change makes very little difference.

\subsection{Modify acceleration table for significant speedup}
\label{sec:interpolation}

In order to speed up orbital integration, the orbit code pre-computes a lookup table of acceleration values over a spatial grid and performs a trilinear interpolation to closely approximate the true acceleration.
If an orbit passes outside the radial range of this grid, the acceleration is then computed from scratch, which is multiple orders of magnitude slower than interpolating values from the lookup table. It is therefore prudent to choose the extent of the grid wisely because even a small number of orbits passing outside the table's coverage can dominate the total runtime and unnecessarily increase the computation time of the entire orbit library.  

We have noticed that some orbits can indeed pass outside the radial range used in the original code and result in a significant slow down. To eliminate this situation,  we have made a simple modification to the radial range used for the acceleration table.
In \citet{vandenBoschetal2008}, the acceleration is pre-computed over a grid spanning the radial range
\begin{eqnarray}
      r_\mathrm{interp,min} & = & \min{\left[ 0.1\times \min{(\sigma'_i)}, 0.01\, r_\mathrm{min} \right]} \nonumber\,,  \\
      r_\mathrm{interp,max} & = &  \max{ \left[6\times \max{(\sigma_i)}, 1.05\, r_\mathrm{max} \right]} \,,
      \label{eq:rminmax}
\end{eqnarray}
where $\sigma'_i$ is length of the semi-major axis of the $i$th projected MGE component, $\sigma_i$ is the length of the semi-major axis of the corresponding intrinsic MGE component, and $r_\mathrm{min}$ and $r_\mathrm{max}$ are the innermost and outermost orbital equipotential radii in the model. Thus, the lowest and highest energy orbits included in the model have energies $\Phi(x=r_\mathrm{min},y=0,z=0)$ and $\Phi(x=r_\mathrm{max},y=0,z=0)$, where $\Phi$ is the gravitational potential of the model.  

\begin{figure}
  \centering
  \includegraphics[width=\linewidth]{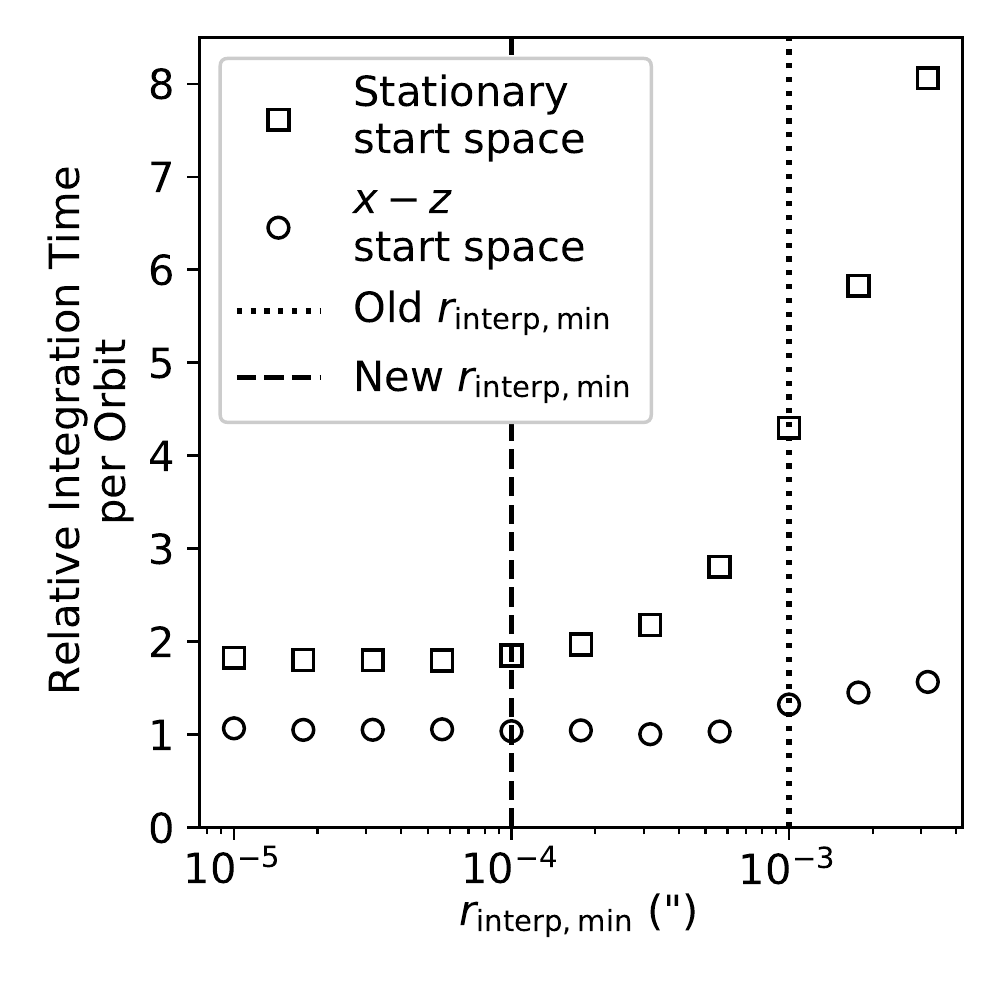}
  \hfill
  \caption{ Average orbital integration time (per orbit) as a function of the inner interpolation radius, $r_\mathrm{interp,min}$, used to tabulate the accelerations. The stationary start space contains mostly box orbits that pass near the galaxy center. The box orbit integration time increases drastically with $r_\mathrm{interp,min}$, and the value used in the \citet{vandenBoschetal2008} code is typically not small enough to minimize the integration time.
  \label{figure:interpolation}}
\end{figure}

In practice, we find that the second conditions in Equation~(\ref{eq:rminmax}) typically determine the range of the acceleration table, i.e., $r_\mathrm{interp,min}=0.01\, r_\mathrm{min}$ and $r_\mathrm{interp,max} = 1.05\,r_\mathrm{max}$. 
The outer boundary is never exceeded because energy conservation prevents orbits from passing outside $r_\mathrm{max}$ and therefore $r_\mathrm{interp,max}$.  The inner boundary of $r_\mathrm{interp,min}=0.01\,r_{\rm min}$, however, can be problematic because centrophilic box orbits can pass well within $0.01\, r_{\rm min}$. The \texttt{DOP853} Runga-Kutta integrator in the TriOS code uses adaptive timesteps, tuning them to minimize errors in the position and velocity between timesteps. In this scheme many acceleration evaluations are required in regions of the trajectory where the timestep is smaller, namely, when the trajectory passes closest to the central black hole where the orbits are most likely to reach below $0.01\, r_{\rm min}$. The fraction of acceleration evaluations within this boundary is somewhat model-dependent and may be higher when box orbits are launched from well within the SMBH's sphere of influence because the potential felt by those orbits is largely spherical and supportive of highly centrophilic box orbits.  For a typical case of $r_\mathrm{min} = 0.1''$ and $r_\mathrm{interp,min} = 0.01\, r_{\rm min} = 0.001''$, we find as many as a sixth of the acceleration evaluations during the box orbit integrations to lie outside the lookup table. This minority of acceleration evaluations take up more than 50\% of the total time when constructing the orbit library.

To enable a more efficient use of the acceleration table, we choose to decouple $r_\mathrm{interp,min}$ and $r_\mathrm{interp,max}$ from $r_\mathrm{min}$ and $r_\mathrm{max}$  which are used to determine the range of orbital energy sampling.  When $r_\mathrm{interp,min}$ is allowed to be smaller than $0.01\, r_{\rm min}$, we find the total time to integrate orbits can be reduced by a factor of a few, with a negligible change in accuracy.
This speed-up is illustrated in Figure~\ref{figure:interpolation}. As the acceleration table is extended to smaller radii,
fewer orbits fall outside the radial coverage of the table, and the average integration time for box orbits drops significantly with decreasing $r_\mathrm{interp,min}$.
For the example shown in Figure~\ref{figure:interpolation}, choosing $r_\mathrm{interp,min}\sim 0.0001''$ would reduce
the orbit integration time by a factor of $\ga 2$ compared with the 
original setting of $r_\mathrm{interp,min}=0.01\, r_{\rm min} = 0.001''$. Since energy conservation prevents orbits from passing outside $r_\mathrm{max}$, setting $r_\mathrm{interp,max}$ to be slightly larger than $r_\mathrm{max}$ minimizes the integration time while maximizing the interpolation accuracy.

Since we don't typically vary the interpolation boundaries by more than 1 dex, the density of points in the interpolation grid does not change dramatically, and we find that the accuracy of the interpolated potential is sufficient. However, if the boundaries are changed more drastically, the number of radial interpolation points should be adjusted to maintain the desired accuracy.

\begin{figure}
  \centering
  \includegraphics[width=\linewidth]{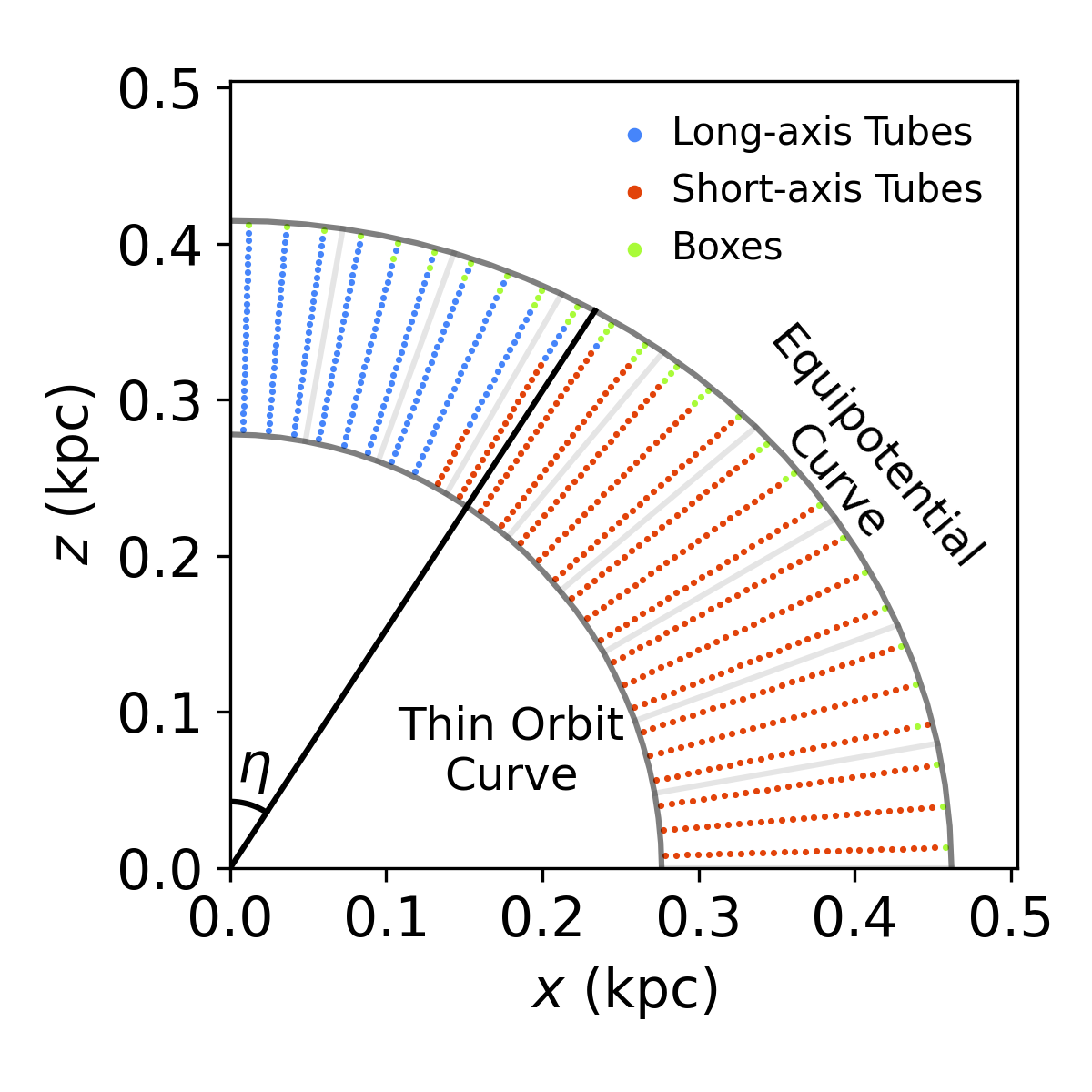}
  \hfill

  \caption{An example of the initial orbit locations in the $x-z$ start space for a single energy value in the triaxial TriOS code.  Orbits are launched from within the thin-orbit curve (inner grey arc) and equipotential curve (outer grey arc). The orbit initial conditions are sampled with $N_\mathrm{I_2}=9$ radial rays uniformly spaced in the polar angle from the $z$-axis to the $x$-axis, $N_\mathrm{I_3}=9$ points along each ray, and $N_{\rm dither}=3$ to further improve the sampling, resulting in a total of 
   $27\times 27$ orbits.
  Each of the $27\times27$ color dots indicates the initial locations of an orbit (color coded by the type of orbits). 
  The black line at angle $\eta$ (see text) approximates the boundary between long-axis and short-axis tube orbits within this start space. Model $\chi^2$ values are sensitive to the alignment between the angle $\eta$ and orbit cell boundaries.
  \label{figure:startspace}}
\end{figure}

\subsection{Resolve issues with insufficient orbit sampling}
\label{sec:1453_sampling}

The TriOS code samples orbit initial conditions from two separate spaces, referred to as start spaces \citep{Schwarzschild1993, vandenBoschetal2008}. 
In the first start space (``stationary start space"), all orbits start from rest on the equipotential surface for a given energy. This start space contains only box, box-like, and chaotic orbits.

The second start space (``$x-z$ start space") contains mainly tube orbits and samples orbits in the $x-z$ plane, with velocity vectors pointing along the $y$-axis. As illustrated in Figure~\ref{figure:startspace}, orbits of a given energy in this space are sampled over the region bounded by the equipotential and thin-orbit curves. 
Typically, $N_\mathrm{I_2}=9$ rays of orbits are sampled uniformly in polar angles from $0$ to $\pi/2$ in the positive $x$ and $z$ quadrant; 
along each ray, $N_\mathrm{I_3}=9$ orbits are uniformly spaced between thin-orbit curves and equipotential curve. Additionally, the code allows for dithering, where orbits with $N_{\rm dither}$ adjacent initial conditions in each dimension are integrated and then bundled together to form each of the $9\times9$ orbits in order to improve phase space sampling. Figure~\ref{figure:startspace} illustrates the case of $(N_\mathrm{I_2}, N_\mathrm{I_3},N_{\rm dither})=(9,9,3)$,  where $27\times 27$ tube orbits are launched in the positive quadrant of the $x-z$ start space for a given energy.

For a triaxial model, 
the short-axis tubes (red points in Figure~\ref{figure:startspace}) and long-axis tubes (blue points) occupy two regions of the $x-z$ start space 
separated by the focal curve. As derived in Appendix A of \citet{Quennevilleetal2021}, the focal curve is roughly approximated by a line at angle
\begin{equation}
    \eta=\tan^{-1} \sqrt{\frac{T}{1-T}}\,.
\end{equation}
Thus, as $T$ increases from 0 to 1, the focal curve moves smoothly from the $z$-axis to the $x$-axis, and the composition of the tube orbits changes from being all short-axis tubes (for an oblate axisymmetric potential) to all long-axis tubes (for a prolate axisymmetric potential).

When orbits are well sampled, model properties such as the goodness-of-fit ($\chi^2$) should vary smoothly as $\eta$ (and hence $T$) is varied.  In our test runs for NGC 1453, however,  we find that on top of a smooth variation, $\chi^2$ varies periodically with $T$ with a frequency matching the spacing between dithered orbits, $(\pi/2)/N_\mathrm{I_2}$, resulting in 
multiple spurious local minima at different values of $T$. Further testing reveals that these local minima arise from insufficient orbit sampling:
as $T$ increases, the focal curve approximated with $\eta$ crosses rays of orbits in a periodic manner, resulting in the artificial oscillations in $\chi^2$ with that same period.
Since the periodic behavior is 
coherent as other model parameters are changed, it can have a significant impact on the recovered value of $T$ and its uncertainty. Other parameter values are mainly impacted through their correlations with $T$.

We are able to eliminate the spurious oscillations in $\chi^2$ vs. $T$ by increasing $N_{I_2}$, which increases the number of radial rays in the $x-z$ start space and therefore improves the sampling in the polar angle.
For the models presented in Section~\ref{sec:dyn_models}, we find that increasing $N_{I_2}$ from the default value of 9 to 15 and beyond removes the oscillations and also yields convergent results.  We choose $N_{I_2}=18$ for the $x-z$ start space.

We do not find similar issues for the other start space.
Nonetheless, we increase $N_{I_2}$ to 18 for the stationary start space as well so as to maintain equal sizes for the tube and box orbit libraries. In summary, we use $(N_E, N_{I_2}, N_{I_3}, N_\mathrm{dither})=(40, 18, 9, 3)$ for both start spaces.  This results in a total $40\times 18 \times 9 \times 3^3 \times 3 = 524,880$ integrated orbits in each galaxy model, where the last factor of 3 accounts for the 3 orbit libraries (the $x-z$ start space, its time-reversed copy, and the stationary start space).

\begin{figure}
  \centering
  \includegraphics[width=3.5in]{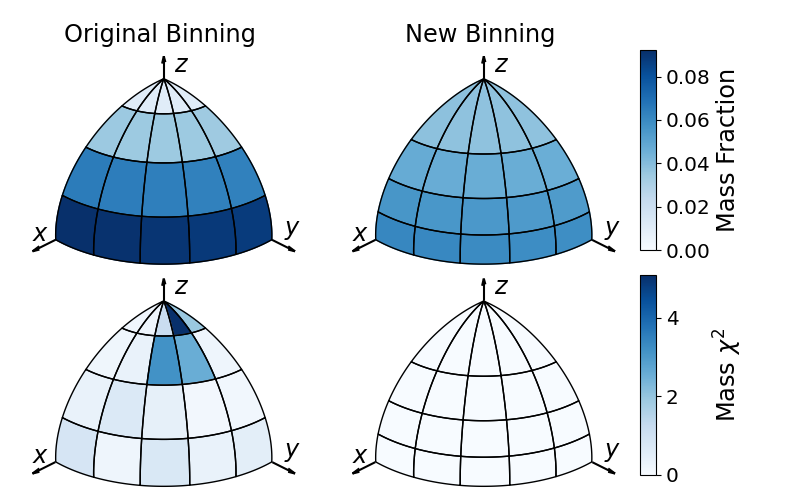}
  \hfill

  \caption{ Comparison of the original (left) and new (right) mass binning scheme in the TriOS code. 
    The top row shows that the bins near the $x-y$ plane contain far more mass than the bins near the $z$ axis due to the significant difference in bin volume in the original scheme (top left). Our new binning scheme evens out the mass considerably (top right).
  The color scale here indicates the fraction of mass that falls within a given angular bin, summed over radius. The bottom row shows an example of the resulting $\chi^2$ in the mass fits for a triaxial galaxy for the two binning schemes. The color scale here indicates $\chi^2$ from attempting to fit a particular mass model, summed over radius. Only the 3D mass distribution is fit, with an error of 1\% assumed on each bin. The most significant contributions to the mass $\chi^2$ are from bins near the $z$-axis that contain very little mass. The triaxial mass model shown here has $\mbh=2.9\times10^9 M_\odot$, $\ml=2.0$, $T=0.10$, $q=0.96q'$, $\Tmaj = 0.95$, and $\Tmin = 0.12$. }
  \label{figure:massgrid}
\end{figure}

\subsection{Improve intrinsic mass binning scheme}
\label{sec:massgrid}

In addition to kinematic constraints, the TriOS code enforces self-consistency of the mass model by requiring that the orbital weights be chosen to reproduce an input mass distribution (e.g., deprojected surface brightness profile of a galaxy). This is done by binning the mass in spherical coordinates $(r,\theta,\phi)$, and requiring that the mass in each bin be reproduced to within a pre-specified precision (typically 1\%).
\citet{vandenBoschetal2008} uses linearly spaced bins between $0^\circ$ and $90^\circ$ for $\theta$ and $\phi$, and logarithmically spaced bins between $r_{\rm min}$ and $r_{\rm max}/2$ for $r$, 
where $r_{\rm min}$ and $r_{\rm max}$ are the innermost and outermost equipotential radii discussed in Section~\ref{sec:interpolation}.

In the axisymmetrized TriOS code \citep{Quennevilleetal2021}, we changed the radial binning scheme above to ensure sufficient orbits are used to represent the innermost and outermost mass bins.
During our subsequent tests for triaxial systems, however, we noticed occasional problems with mass misfits in which a handful mass bins would have difficulty satisfying the 1\% precision and/or contribute disproportionately high values to the total $\chi^2$ of the galaxy model under examination.  We are able to trace the problem to uneven bin sizes in $\theta$ used in the original code: the bins near the poles contained much less mass, as shown in the left panel of Figure~\ref{figure:massgrid}.  Because of this, the mass near the $z$ axis was subject to much more stringent constraints than elsewhere, leading to frequent difficulties in satisfying the 1\% fitting criterion.
Even in the absence of kinematic constraints, spurious variations would arise in the $\chi^2$ landscape, as illustrated in the right panel of Figure~\ref{figure:massgrid}.  The more oblate ($T\lesssim0.1$) and round ($q\gtrsim0.9q'$) systems
are more prone to this issue.

We find that this mass misfitting problem can be easily resolved by using mass bins linearly spaced in $\cos{(\theta)}$ and $\phi$, rather than in $\theta$ and $\phi$.  The resulting bins at a given radius
then occupy the same volume, and the mass in each bin is much more uniform, with the bin-to-bin variations representing the galaxy's intrinsic deviation from spherical symmetry. Correspondingly, the pre-specified mass constraint criterion is enforced more uniformly.

For clarity, we have chosen to illustrate the mass misfitting issue in Figure~\ref{figure:massgrid} without imposing any kinematic constraints.  When kinematic constraints are added in full orbit modeling (see Section~\ref{sec:dyn_models}),
the total $\chi^2$ returned by the code includes
contributions from fits to the masses as well as kinematics.  In this case, models with significant mass misfits due to uneven binning schemes would have disproportionately larger $\chi^2$ values, leading to potential biases in the recovered galaxy parameters.

\section{Triaxial Orbit Models of NGC~1453}
\label{sec:dyn_models}

\subsection{NGC~1453}

We apply the updated TriOS code described in the previous section to NGC~1453, a massive elliptical galaxy targeted by the MASSIVE survey \citep{Maetal2014}.
In \citet{Liepoldetal2020}, we performed orbit modeling of NGC~1453 using the axisymmetrized TriOS code.
We refer the reader to that paper for a detailed description of the input kinematic and photometric data.  In brief, the stellar kinematics are measured over 135 spatial bins from our high-spatial resolution Gemini GMOS IFS data \citep{Eneetal2019, Eneetal2020} and wide-field McDonald Mitchell IFS data \citep{Vealeetal2017a, Vealeetal2017b, Vealeetal2018}. 
The first eight Gauss-Hermite moments are measured from the IFS spectra and used to constrain the stellar LOSVD in each kinematic bin; see Figure~4 of \citet{Liepoldetal2020}.

The MGE components representing the galaxy's mass distribution (see Section~\ref{sec:deprojection}) are obtained from deprojections of our \emph{HST} WFC3 photometry \citep{Goullaudetal2018}. Here we use the same input data but relax the assumption of axisymmetry in the orbit models. In order to ensure that all trajectories within the model are representative of their equilibrium distributions, we integrate each orbit in the $x-z$ start space for 2000 times the orbital period for a thin tube orbit of the same energy. For orbits in the stationary start space, we integrate for 200 times the orbital period, as is typical of previous studies using the \citet{vandenBoschetal2008} code.

\begin{figure*}
  \centering
  \includegraphics[width=6in]{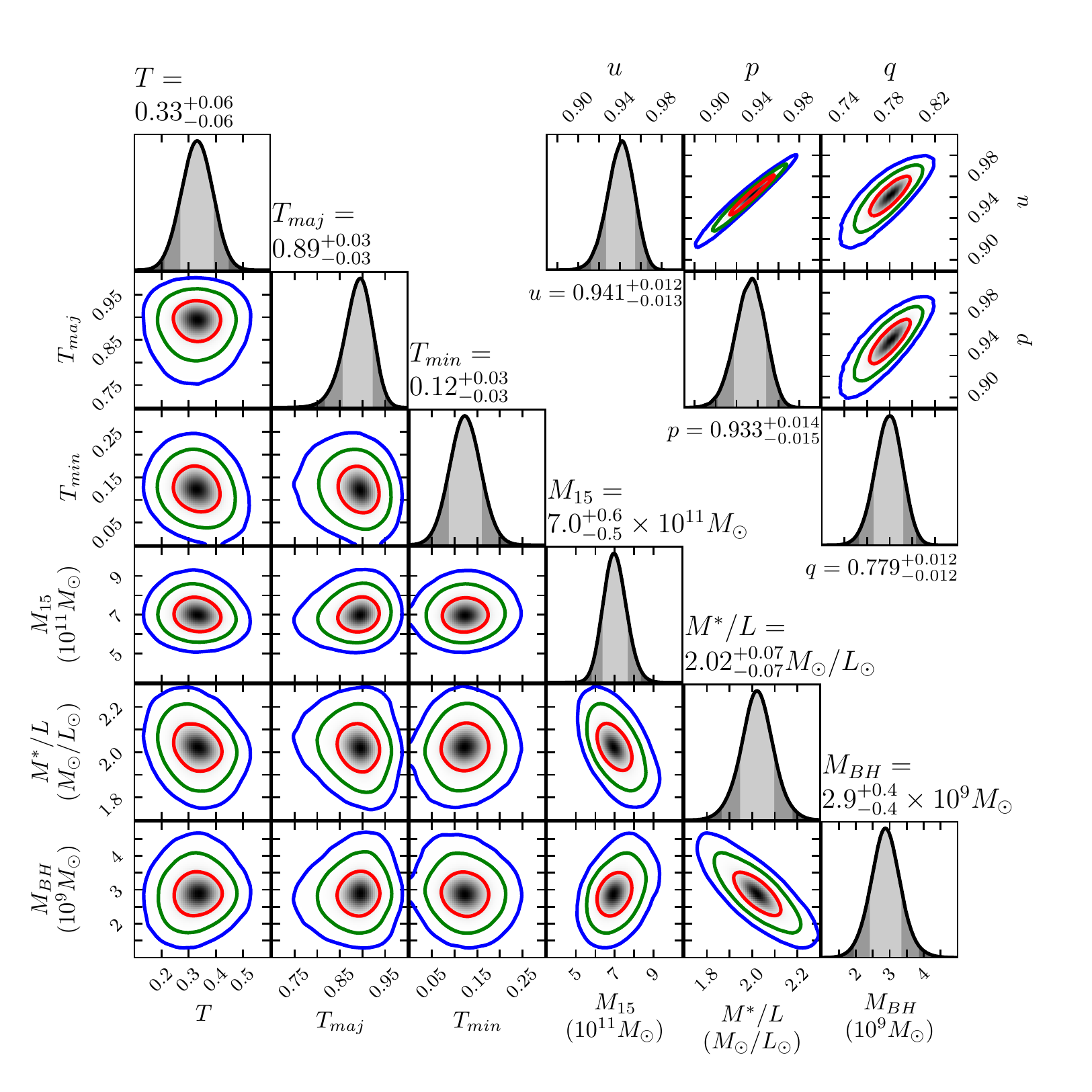}
  \hfill
  \caption{(Left) 5D likelihood landscape for orbit models of NGC 1453. As described in the text, the models are sampled in $T$, $\sqrt{\Tmaj}$, $\sqrt{\Tmin}$, $\mbh$, $\ml$, and $\mdm$, and the 1D and 2D likelihood landscapes are obtained by marginalizing over a smoothed 5D landscape generated by Gaussian process regression.  The red, green, and blue curves represent the $1\sigma$, $2\sigma$, and $3\sigma$ contours, respectively. (Right) 3D likelihood in axis ratio space, $(p, q, u)$, marginalized over $\mbh$, $\ml$, and $\mdm$.  
  All three axis ratios are significantly correlated with one another, in particular between $p$ and $u$.  This degeneracy is significantly reduced when our new shape parameters $T, \Tmaj$, and \Tmin\ are used.}
  \label{figure:5d_shape}
\end{figure*}

Due to the regular isophotes of NGC 1453 (Figure~5 of \citealt{Liepoldetal2020}), we use the same
PA for all MGE components and do not model isophotal twists. This is a common simplifying assumption \citep[e.g.,][]{vandenBoschdeZeeuw2010,Walshetal2012,FeldmeierKrause2017} and
it enables us to explore the galaxy's shape using the new scheme outlined in Section~\ref{sec:new_params}. 

For the distance to NGC~1453, we adopt our new determination of 51.0 Mpc from the MASSIVE-WFC3 project \citep{Goullaudetal2018} using the surface-brightness fluctuation technique \citep{Jensenetal2021}.  
At this distance, $1''$ is 245 pc for a flat $\Lambda$CDM model with a matter density of $\Omega_m = 0.315$ and a Hubble parameter of $H_0=70 \kms$ Mpc$^{-1}$.

\subsection{Parameter Search Using Latin Hypercube Sampling}
\label{sec:1453_shape}

We conduct the search for the best-fit galaxy shape in the new triaxial parameters $(T, \Tmaj, \Tmin)$ introduced in Section~\ref{sec:new_params}. The dark matter halo is modelled as a logarithmic potential. We parameterize it through its mass within 15 kpc, $\mdm$, which is roughly the central radius of the outermost kinematic bins, following \citet{Liepoldetal2020}. As in \citet{Liepoldetal2020}, we fix the scale radius of the dark matter halo to 15 kpc. Combining the three shape parameters with the three mass parameters $\mbh$, $\ml$, and $\mdm$, we sample the 6D parameter space of galaxy models. 

We determine the best-fit parameters by 
minimizing a $\chi^2$ that includes terms for each 
LOSVD moment within each aperture, the projected light within each aperture, as well as the binned 3D mass density in order to enforce self-consistency for the stellar density. For each model, the best-fit set of weights are used to calculate the $\chi^2$ differences between models. \citet{LipkaThomas2021} recently suggested that recovery of the inclination of axisymmetric models can be biased unless the intrinsic flexibility of the models is accounted for. However, a triaxial exploration of model flexibility is beyond the scope of the present study.

Instead of conducting model searches on a regular grid as was done in previous studies, we use the more efficient method of Latin hypercube sampling \citep{McKayetal1979}.  There are many techniques for ensuring spatial uniformity in multidimensional spaces. We adopt the scheme described in \citet{deutschdeutsch2012}, as implemented in the \texttt{LHSMDU} python package \citep{LHSMDU}.
This procedure results in models that span a more continuous range of values than a regular grid, and are more uniformly spaced than random sampling.
This approach allows a more representative sampling of the 6 dimensions with many fewer points than a regular grid.

We initially use a hypercube consisting of 1000 models spanning the range of $\ml \in [1.7,2.3]$, $\mdm \in [3.5,10.5]\times10^{11} M_\odot$, and $\mbh \in [1,5]\times10^9 M_\odot$,
and the full range between 0 and 1 for $(T,\sqrt{\Tmaj},\sqrt{\Tmin})$.
Of these models, 927 resulted in valid deprojections.
We then use a rejection-based scheme to choose subsequent sets of model points. A Gaussian process interpolation of the 6-dimensional $\chi^2$ surface is computed from the previously-run models. We use this interpolation to estimate the $\chi^2$ for $O(10^4)$ points chosen using the LHS scheme described above in the original volume and select points where the estimated $\chi^2$ is within $\Delta \chi^2 = 20.06$ ($3\sigma$ for 6 parameters) of the estimated global minimum. To avoid premature optimization we perform this routine 10 times where random subsets of half of all previously-run models are used to build the interpolation function. With this scheme we select roughly 1000 model points which are expected to lie near the global $\chi^2$ minimum to evaluate with the TriOS code. We perform two iterations of this rejection scheme, yielding roughly 3000 total model evaluations.

The resulting 6D likelihood landscape is shown in Figure~\ref{figure:5d_shape}.  To determine the best-fit value and uncertainties, we fit the $\chi^2$ landscape using Gaussian process regression with a squared-exponential covariance function \citep{scikit-learn}. To make the 2D contours shown in Figure~\ref{figure:5d_shape}, 
we transform this smoothed surface from $(T, \sqrt{\Tmaj}, \sqrt{\Tmin})$ to $(T, \Tmaj, \Tmin)$, or $(p, q, u)$.  The marginalized 1D likelihood is also shown for each parameter. The shapes of the 2D contours in Figure~\ref{figure:5d_shape} clearly demonstrate that $(T, \Tmaj, \Tmin)$ do not have the strong degeneracy apparent in $(p, q, u)$.

The standard values of $\Delta \chi^2=1,4,9$ are used to define the $1\sigma$, $2\sigma$, and $3\sigma$ confidence intervals for 1 degree of freedom when considering the marginalized landscape for each variable individually. For the 2D contours, we use the values for 2 degrees of freedom, giving $\Delta \chi^2\approx2.3,6.2,11.8$. This is different from most previous work using the \citet{vandenBoschetal2008} code, where typically $\Delta\chi^2=\sqrt{2N_\mathrm{obs}}$ is used to define the $1\sigma$ confidence interval, where $N_\mathrm{obs}$ is the number of apertures on the sky, multiplied by the number of moments fitted within each aperture. This value is chosen to represent the intrinsic noise in the $\chi^2$ values for each model, and is much larger than our values. However, while this is true when the input data are varied according to its noise level as discussed in \citet{VasilievValluri2020}, the noise level in the $\chi^2$ values between models are significantly smaller when the input data are fixed. 

\subsection{Best-fit Triaxial Model}

The best-fit values and the uncertainties for each \ngc{} parameter are listed in Table~\ref{tab:parameters}. For each parameter, all other dimensions have been marginalized over. The best-fit \mbh{} is consistent with the value determined from axisymmetric modelling in \citet{Liepoldetal2020}. The value of \ml{} has shifted down slightly, but is still consistent within $2\sigma$ of the axisymmetric value.  

The best-fit shape, on the other hand, is inconsistent with axisymmetry. It is useful to compare our best-fit values of $p=0.93$ and $q=0.78$ with those
 inferred statistically from
the observed distributions of ellipticity  and misalignment angle between the kinematic and photometric axes for 49 slowly-rotating massive elliptical galaxies with measurable kinematic axes in the MASSIVE survey \citep{Eneetal2018}.
In that sample, 56\% of the galaxies have $p > 0.9$ with a mean value of 0.88, and the mean value of $q$ is 0.65.
Our best-fit shape for NGC~1453 indicates this fast-rotating galaxy is relatively oblate like the MASSIVE slow rotators and is slightly less flattened than the mean of that population.

\begin{table}[]
\begin{tabular}{c|c}
Parameter                & Value \\ \hline
$\mbh\:(10^9 M_\odot)$    & $2.9\pm0.4$ \\
$\ml\:(M_\odot/L_\odot)$  & $2.02\pm0.07$ \\
$\mdm\:(10^{11} M_\odot)$    & $7.0^{+0.6}_{-0.5}$ \\
$T$                      & $0.33\pm0.06$ \\
$\Tmaj$ & $0.89\pm0.03$ \\
$\Tmin$ & $0.12\pm0.03$ \\
$u$                      & $0.941^{+0.012}_{-0.013}$      \\
$p$                      & $0.933^{+0.014}_{-0.015}$     \\
$q$                      & $0.779\pm0.012$      \\
$\theta\:(^\circ)$        & $73\pm3$      \\
$\phi\:(^\circ)$                   & $19\pm3$      \\
$\psi\:(^\circ)$                   & $92.7^{+0.7}_{-0.8}$
\end{tabular}
\caption{ Best-fit triaxial model parameters for NGC~1453 from the 6D likelihood landscape in Figure~\ref{figure:5d_shape}. For each parameter, all other dimensions have been marginalized over.
\label{tab:parameters}}
\end{table}

\begin{figure}
\centering
  \includegraphics[width=\linewidth]{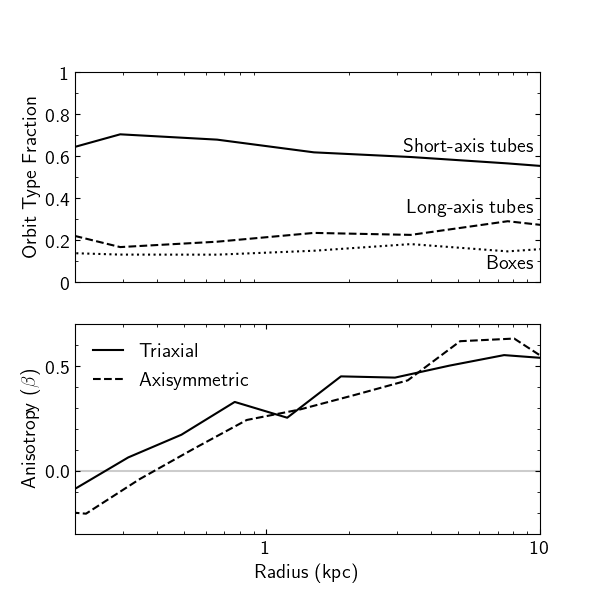}
  \hfill
  \caption{Orbital composition (top) and velocity anisotropy (bottom) of the best-fit triaxial model of NGC~1453 as a function of radius. 
  Short-axis tubes (solid) are dominant throughout the model, with significant contributions from long-axis tubes (dashed) and box orbits (dotted) that are present only in triaxial potentials. 
  The velocity anisotropy parameter, $\beta$, has a similar radial profile for the best-fit triaxial (solid) and axisymmetric (dashed) models, being mildly tangentially anisotropic in the inner part and becoming more radially anisotropic in the outer part.
  \label{figure:profiles}}
\end{figure}

The orbital composition of the best-fit triaxial model is shown in Figure~\ref{figure:profiles} (top panel). Long-axis tubes and box orbits -- two orbit types that are present only in triaxial potentials -- together account for $\sim 30$\% of the orbital weights in the inner part and $\sim 45$\% in the outer part of NGC~1453.
Quasi-planar orbits account for a small fraction of the total mass at small and large radii and are excluded from the plot. While long-axis tubes contribute a significant fraction of the mass, the projected model has fairly little minor axis rotation, due in part to the LOS being close to the intrinsic major axis. 

The orbital velocity anisotropy of the best-fit model (bottom panel of Figure~\ref{figure:profiles}) is mildly tangential ($\beta < 0$) in the inner part and becomes increasingly radial outward.
The radial profile has a similar shape to the axisymmetric model presented in \citet{Liepoldetal2020}.

\subsection{Triaxial vs. Axisymmetric Best-fit Models}
\label{sec:triaxial_axisym}

\begin{figure*}
  \centering
  \includegraphics[width=\linewidth]{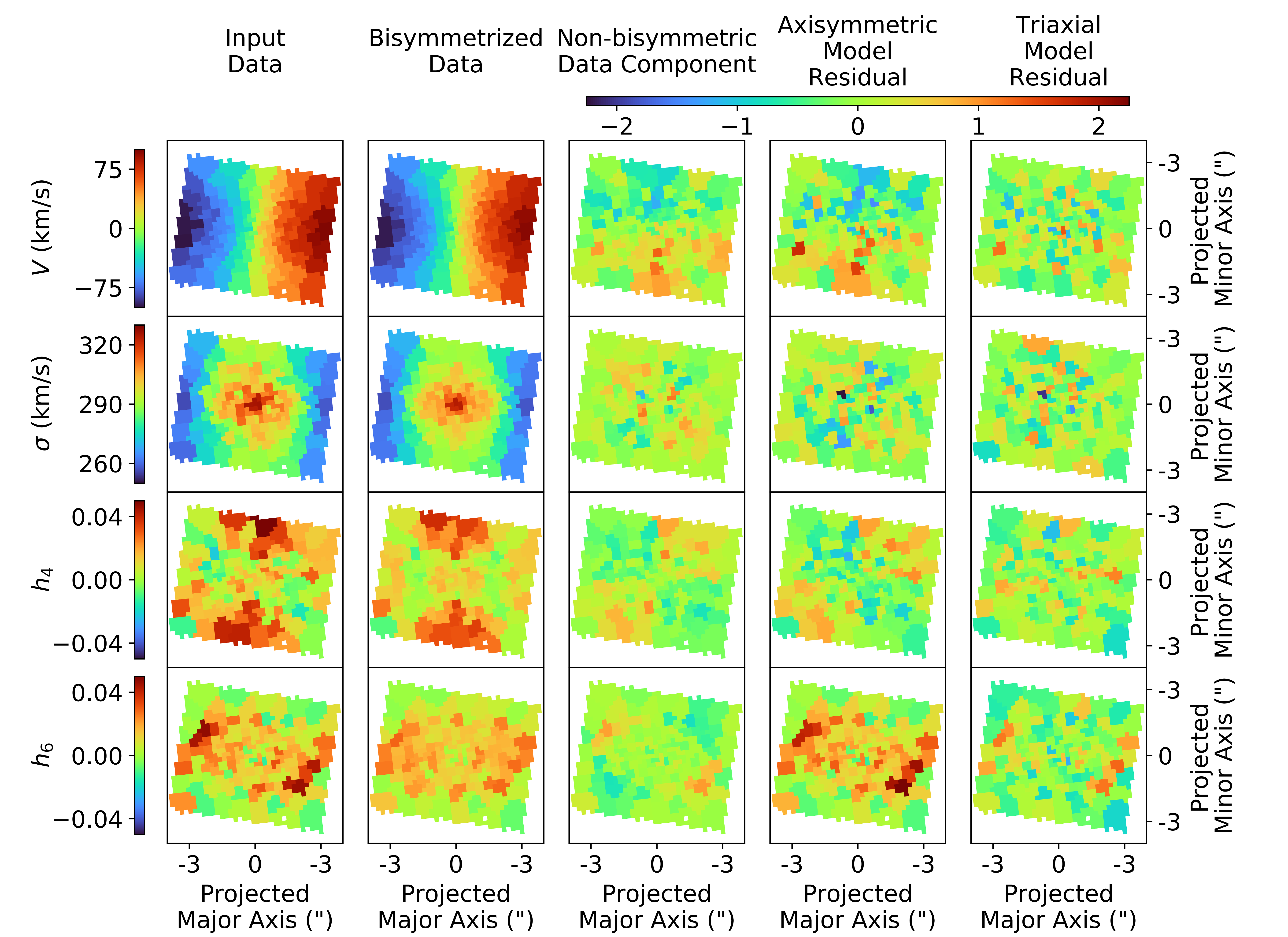}
  \hfill

  \caption{Maps of the stellar kinematics from the Gemini GMOS IFS in 135 spatial bins of the central $5''\times 7''$ of NGC~1453. Four velocity moments are shown (from top down): $V$, $\sigma$, 
  $h_4$ and $h_6$. The maps are oriented such that the horizontal and vertical axes are aligned with the galaxy's projected major and minor photometric axes, respectively. The data (first column) are decomposed into a bisymmetric component (second column) and a non-bisymmetric component (third column).
  To accentuate systematic patterns, we plot the non-bisymmetric component normalized by the moment uncertainty.  Since an axisymmetric model can only produce bisymmetric kinematic maps, the residuals from the best-fit axisymmetric model (fourth column) show similar patterns to the bisymmetrized residuals. $h_6$ shows additional residuals that are consistent with bisymmetry, but unable to be fit by an axisymmetric model. A triaxial model (right column) is able to capture most of the systematic behaviour in the input map, resulting in largely random residuals. The residuals have been normalized by the moment uncertainty.
  \label{figure:maps}}
\end{figure*}

The best-fit triaxial model presented above matches the observed kinematics significantly better than the best-fit axisymmetric model in \citet{Liepoldetal2020}.

Even though the best-fit $\chi^2$ values in the two cases -- 493.0 for axisymmetric versus 382.7 for triaxial -- differ by $\sim110$, they should not be compared directly because triaxial potentials require  a  new  library  of  box  orbits, and different numbers of orbits are used (6480 independent weights for axisymmetric versus 19440 for triaxial). 
Nonetheless, within triaxial modeling, our best-fit triaxiality of $T=0.33$ is preferred over nearly oblate axisymmetric models with $T\approx 0$ at a confidence level of about 5$\sigma$.
 To understand why non-axisymmetric models are favored, we examine the 2D maps of $V$ and the lowest 3 even Gauss-Hermite moments in the GMOS data in Figure~\ref{figure:maps} (first row).
We recall that axisymmetric models by construction produce only bisymmetric kinematics about the photometric major axis on the sky, meaning that the LOSVDs would be symmetric for points mirrored across the projected major axis and anti-symmetric for points mirrored across the projected minor axis.
Any observed systemic deviation from bisymmetry would then indicate triaxiality.

For this reason,
we decompose each GMOS moment map into a bisymmetrized component (second column) and a non-bisymmetrized component (third column).
The latter exhibits clear systemic deviations from bisymmetry. The most obvious feature is the residual minor axis rotation indicative of kinematic misalignment. These maps assume a bisymmetrization along the projected photometric major axis used by our dynamical models, with a PA of $28.5^\circ$. The residual pattern persists and can not be ``rotated away" even if the PA is within uncertainties in the PA determination determined from the isophotal profile from \citep{Goullaudetal2018}.
An axisymmetric model (consistent with the photometry) would be incapable of fitting these non-bisymmetric features in the data.
To confirm this point, we plot the residual maps (fourth column) between the GMOS data and the best-fit axisymmetric model of \citet{Liepoldetal2020}. Indeed, the axisymmetric model exhibits similar residual patterns as in the data (third column). 
In comparison, the best-fit triaxial model is able to fit these non-bisymmetric features to a large extent, producing essentially random residuals (fifth column).

Figure~\ref{figure:maps} indicates that
the preference for triaxiality is driven by the non-axisymmetric features in the NGC~1453 kinematics.
Even though the non-bisymmetric features are somewhat subtle, they lead to detectable triaxiality, which we find to be best fit with $p=0.933$, $q=0.779$, and $T=0.33$. Thus, despite being a fast rotator with regular isophotal and kinematic features, NGC~1453 is best fit by a triaxial model. This is further evidence for widespread triaxiality in massive elliptical galaxies. 

 Importantly, however, the best-fit black hole mass $\mbh = 2.9\times 10^9 M_\odot$ is unchanged from that in the axisymmetric model. The stellar mass-to-light ratio and dark matter mass within 15 kpc agree to within a $1\sigma$ confidence level.

\section{Conclusions}
\label{sec:conclusion}

In this paper we have presented a revised code and a revamped approach for performing dynamical modeling of triaxial galaxies and their central SMBHs using the orbit superposition technique. We discussed a new triaxial version of the TriOS code that is capable of modeling triaxial systems while avoiding several shortcomings of the original \citet{vandenBoschetal2008} code.  As a first application of this code, we performed triaxial orbit modeling of the massive elliptical galaxy NGC~1453
and presented the best-fit galaxy shape and mass parameters.  This work complements \citet{Liepoldetal2020} and \citet{Quennevilleetal2021}, in which we introduced a properly axisymmetrized version of the TriOS code.  

We discovered and corrected a major error in the orbit kinematics in the \citet{vandenBoschetal2008} code: the tube orbits had wrong signs in certain mirrored velocity components in the orbit library (Table~1), resulting in incorrect projected kinematics. The magnitude of the kinematic errors varies spatially and depends on the viewing angles (Figure~\ref{figure:flips}). This issue impacts all triaxial models that are not viewed along a principal axis, and all nearly axisymmetric models that are not viewed edge-on.  How this error affects the best-fit galaxy shapes and mass parameters would have to be assessed on a galaxy-by-galaxy basis by re-running the models with the corrected orbital flips in Table~1. In the case of \ngc{}, we find the $\chi^2$ landscapes to be altered drastically, with $\chi^2$ values changing non-uniformly by more than 100 for some models. 

Following \citet{Quennevilleetal2021}, we continued to find ways to speed up the code.
In this updated version of the TriOS code,
we achieved another significant speedup (of up to $\sim 50~\%$; Figure~\ref{figure:interpolation}) in orbit integration time by a simple extension of the interpolation table used to evaluate orbit accelerations (Section~\ref{sec:interpolation}). The reduction in integration time is particularly pronounced for centrophilic orbits. 

We have made two other adjustments in the code
that significantly improve the sampling of long-axis tube orbits  (Section~\ref{sec:1453_sampling}) and enforce more uniformly the 3D mass constraints (Section~\ref{sec:massgrid}). 
After these changes, the behavior of $\chi^2$ vs. $T$ (triaxiality parameter) no longer exhibits spurious oscillations, and the orbit code is able to find reasonable solutions for some mass models that were previously strongly disfavored. 

The rest of this paper is devoted to new and improved strategies for searching the multi-dimensional parameter space required to specify triaxial galaxy models. We introduced a new set of shape parameters (Section~3) as well as a novel sampling technique (Section~\ref{sec:1453_sampling}), which together lead to a remarkable gain in parameter searching efficiency. Searching in the new parameters $T$, $\Tmaj$, and $\Tmin$ (Equation~\ref{eq:T}) avoids significant non-uniformities associated with other parameters used in earlier work.  Our Latin hypercube sampling scheme results in an order-of-magnitude reduction in needed sampling points compared with conventional grid searches.

We applied the TriOS code and triaxial sampling scheme to the fast-rotating massive elliptical galaxy \ngc{} in the MASSIVE survey (Section~5).  \ngc{} has a relatively small twist in the isophotes, and the kinematic and photometric axes are nearly aligned. Despite these properties that are typically invoked to justify the use of axisymmetric orbit codes, we find the best-fit model to have a triaxiality value of $T=0.33$, with intrinsic axis ratios $p=0.933$ and $q=0.779$.
This best-fit triaxial model is able to match the observed kinematic maps significantly better than the best-fit axisymmetric model in \citet{Liepoldetal2020}. The improvement is mainly due to the ability of triaxial models to account for non-bisymmetric features in the data (Figure~\ref{figure:maps}).
Most other galaxies in the MASSIVE survey exhibit less (or no) rotation and more twists in their photometric and kinematic maps compared to NGC~1453. This is further evidence that massive elliptical galaxies have triaxial intrinsic shapes.

\mbh\ in the best-fit triaxial model for NGC~1453 
is unchanged
from the value measured with the axisymmetrized TriOS code from \citet{Liepoldetal2020}.
Among the many dozens of stellar dynamical \mbh\ measurements in local galaxies (e.g., \citealt{mcconnellma2013}), NGC~1453 is only 
one of a handful galaxies
whose central SMBH is studied with the full triaxial orbit modeling technique not limited to axisymmetry.
In four other galaxies (Section~1),
M32 had consistent \mbh\ from axisymmetric and triaxial modeling, the NGC~3379 \mbh\ increased by a factor of $\sim 2$ when axisymmetry was relaxed, the PGC 046832 \mbh\ decreased enough to be consistent with 0, while NGC~3998 was only modeled with the triaxial code so no comparison can be made.  All four systems were modeled with the original \citealt{vandenBoschetal2008} code, which used the incorrect mirroring scheme.  Triaxial orbit modeling of more galaxies is needed for a full assessment of the systematic effects on stellar dynamical \mbh\ measurements when the commonly-made assumption of axisymmetry is relaxed.  

\begin{acknowledgments}

We thank Shaunak Modak and Jonelle Walsh for useful discussions. M.E.Q. acknowledges the support of the Natural Sciences and Engineering Research Council of Canada (NSERC), PGSD3-517040-2018. C.-P.M. acknowledges support from NSF AST-1817100, HST GO-15265, HST AR-14573, the Heising-Simons Foundation, the Miller Institute for Basic Research in Science, and the Aspen Center for Physics, which is supported by NSF grant PHY-1607611.  This work used the Extreme Science and Engineering Discovery Environment (XSEDE) at the San Diego Supercomputing Center through allocation AST180041, which is supported by NSF grant ACI-1548562. 

\end{acknowledgments}

\appendix
\section{Relating new and old parameters}
\label{sec:ap_shapes}

The expressions given in Equation~(\ref{eq:angles_Ts}) can be written in a simpler form when expressed sequentially:
\begin{equation}
    \label{eq:angles_Ts_seq}
    \begin{split}
    \cos^2{\theta}&=\Tmin(1-T \Tmaj) \,, \\
    \sin^2{\phi}&=\frac{(1 - \Tmaj)(1 - \Tmin)}{\sin^2{\theta}}\,, \\
    \tan{\psi}&=\frac{- (1 - \Tmin) \cos{\theta}}{(\Tmin-\cos^2{\theta})\tan{\phi}}  \,.
    \end{split}
\end{equation}
The inverse expressions are then
\begin{equation}
    \label{eq:angles_Ts_inv}
    \begin{split}
    T&=\frac{\sin^2{\theta}}{\cos{\theta}\sin{2\phi}\cot{2\psi} + \cos^2{\phi} - \cos^2{\theta} \sin^2{\phi}} \,, \\
    \Tmaj&=1-\sin^2{\phi}(1-\cos{\theta}\cot{\phi}\cot{\psi}) \,, \\
    \Tmin&=1-\sin^2{\theta}(1-\cos{\theta}\cot{\phi}\cot{\psi})^{-1} \,.
    \end{split}
\end{equation}

The deprojection equations, giving the intrinsic shape in terms of the projected flattening and angles $(\theta,\phi,\psi)$ are given by \citep{Cappellari2002}:
\begin{equation}
\label{eq:projection_angles}
\begin{split}
    1-q^2&=\frac{\delta'[2\cos{2\psi}+\sin{2\psi}(\sec{\theta}\cot{\phi}-\cos{\theta}\tan{\phi})]}{2\sin^2{\theta}[\delta'\cos{\psi}(\cos{\psi}+\cot{\phi}\sec{\theta}\sin{\psi})-1]} \\
    p^2-q^2&=\frac{\delta'[2\cos{2\psi}+\sin{2\psi}(\cos{\theta}\cot{\phi}-\sec{\theta}\tan{\phi})]}{2\sin^2{\theta}[\delta'\cos{\psi}(\cos{\psi}+\cot{\phi}\sec{\theta}\sin{\psi})-1]} \\
    u^2&=\frac{1}{q'}\sqrt{p^2\cos^2{\theta}+q^2\sin^2{\theta}(p^2\cos^2{\phi}+\sin^2{\phi})},
\end{split}
\end{equation}
where $\delta'=1-q'^2$. While \cite{Cappellari2002} presents these expressions in the context of the MGE formalism, they are more broadly applicable to all densities that are stratified on similar concentric ellipsoids. This is demonstrated in \citet{deZeeuwFranx1989}. The first two expressions in equation \ref{eq:projection_angles} are listed as their equation A8. The third expression giving the projection axis ratio, $u$, follows from expressions in this paper as well. Following appendix A of this paper, combining their equations 3.37, 3.38, and 3.49 gives:
\begin{equation}
\begin{split}
    a'^2+b'^2&=2c^2+(a^2-c^2)(\sin^2{\phi}+\cos^2{\phi}\cos^2{\theta})+(b^2-c^2)(\cos^2{\phi}+\sin^2{\phi}\cos^2{\theta}) \\
    (a'^2-b'^2)^2&=[(a-c^2)(\sin^2{\phi}-\cos^2{\phi}\cos^2{\theta})+(b^2-c^2)(\cos^2{\phi}-\sin^2{\phi}\cos^2{\theta})]^2+4(a^2-b^2)^2\sin^2{\phi}\cos^2{\phi}\cos^2{\theta}.
\end{split}
\end{equation}
Here, $(\alpha,\beta,\gamma)$ in the original expressions have been set to $(-a^2,-b^2,-c^2)$ in order to consider a perfect ellipsoid. The first of these expressions is explicitly given in equation A6 of the original paper. Squaring the first expression and subtracting the second gives (after significant simplification):
\begin{equation}
    4a'^2b'^2=4a^2b^2\cos^2{\theta}+4a^2c^2\sin^2{\phi}\sin^2{\theta}+4b^2c^2\cos^2{\phi}\sin^2{\theta}.
\end{equation}
Substituting the definitions of the axis ratios reduces this expression to the third line of equation \ref{eq:projection_angles} above. 

Equation~\ref{eq:angles_Ts} follows from equations 3.39 and 3.42 of \citet{deZeeuwFranx1989}, together with the definitions given in equation~\ref{eq:shape_params}. Equation~\ref{eq:projection} then follows from equation~\ref{eq:projection_angles}, together with equation~\ref{eq:angles_Ts}. As in appendix A1 of \citet{deZeeuwFranx1989}, while these expressions are derived in the context of a perfect ellipsoid, the results are independent of the assumed profile and are thus valid for all densities stratified on similar concentric ellipsoids.

    \bibliography{tri1453}{}

\begin{thebibliography}{}
\expandafter\ifx\csname natexlab\endcsname\relax\def\natexlab#1{#1}\fi
\providecommand{\url}[1]{\href{#1}{#1}}
\providecommand{\dodoi}[1]{doi:~\href{http://doi.org/#1}{\nolinkurl{#1}}}
\providecommand{\doeprint}[1]{\href{http://ascl.net/#1}{\nolinkurl{http://ascl.net/#1}}}
\providecommand{\doarXiv}[1]{\href{https://arxiv.org/abs/#1}{\nolinkurl{https://arxiv.org/abs/#1}}}

\bibitem[{{Ahn} {et~al.}(2018){Ahn}, {Seth}, {Cappellari}, {Krajnovi{\'c}},
  {Strader}, {Voggel}, {Walsh}, {Bahramian}, {Baumgardt}, {Brodie},
  {Chilingarian}, {Chomiuk}, {den Brok}, {Frank}, {Hilker}, {McDermid},
  {Mieske}, {Neumayer}, {Nguyen}, {Pechetti}, {Romanowsky}, \&
  {Spitler}}]{Ahnetal2018}
{Ahn}, C.~P., {Seth}, A.~C., {Cappellari}, M., {et~al.} 2018, \apj, 858, 102,
  \dodoi{10.3847/1538-4357/aabc57}

\bibitem[{Binney(1985)}]{Binney1985}
Binney, J. 1985, \mnras, 212, 767, \dodoi{10.1093/mnras/212.4.767}

\bibitem[{Cappellari(2002)}]{Cappellari2002}
Cappellari, M. 2002, \mnras, 333, 400, \dodoi{10.1046/j.1365-8711.2002.05412.x}

\bibitem[{{Cappellari}(2016)}]{Cappellari2016}
{Cappellari}, M. 2016, \araa, 54, 597,
  \dodoi{10.1146/annurev-astro-082214-122432}

\bibitem[{{de Nicola} {et~al.}(2020){de Nicola}, {Saglia}, {Thomas}, {Dehnen},
  \& {Bender}}]{deNicolaetal2020}
{de Nicola}, S., {Saglia}, R.~P., {Thomas}, J., {Dehnen}, W., \& {Bender}, R.
  2020, \mnras, 496, 3076, \dodoi{10.1093/mnras/staa1703}

\bibitem[{{de Zeeuw} \& {Franx}(1989)}]{deZeeuwFranx1989}
{de Zeeuw}, T., \& {Franx}, M. 1989, \apj, 343, 617, \dodoi{10.1086/167735}

\bibitem[{{den Brok} {et~al.}(2021){den Brok}, {Krajnovi{\'c}}, {Emsellem},
  {Brinchmann}, \& {Maseda}}]{denBroketal2021}
{den Brok}, M., {Krajnovi{\'c}}, D., {Emsellem}, E., {Brinchmann}, J., \&
  {Maseda}, M. 2021, \mnras, 508, 4786, \dodoi{10.1093/mnras/stab2852}

\bibitem[{Deutsch \& Deutsch(2012)}]{deutschdeutsch2012}
Deutsch, J.~L., \& Deutsch, C.~V. 2012, Journal of Statistical Planning and
  Inference, 142, 763, \dodoi{https://doi.org/10.1016/j.jspi.2011.09.016}

\bibitem[{{Do} {et~al.}(2019){Do}, {Hees}, {Ghez}, {Martinez}, {Chu}, {Jia},
  {Sakai}, {Lu}, {Gautam}, {O'Neil}, {Becklin}, {Morris}, {Matthews},
  {Nishiyama}, {Campbell}, {Chappell}, {Chen}, {Ciurlo}, {Dehghanfar},
  {Gallego-Cano}, {Kerzendorf}, {Lyke}, {Naoz}, {Saida}, {Sch{\"o}del},
  {Takahashi}, {Takamori}, {Witzel}, \& {Wizinowich}}]{Doetal2019}
{Do}, T., {Hees}, A., {Ghez}, A., {et~al.} 2019, Science, 365, 664,
  \dodoi{10.1126/science.aav8137}

\bibitem[{{Emsellem} {et~al.}(2007)}]{emsellemetal2007}
{Emsellem}, E., {et~al.} 2007, \mnras, 379, 401,
  \dodoi{10.1111/j.1365-2966.2007.11752.x}

\bibitem[{{Ene} {et~al.}(2019){Ene}, {Ma}, {McConnell}, {Walsh}, {Kempski},
  {Greene}, {Thomas}, \& {Blakeslee}}]{Eneetal2019}
{Ene}, I., {Ma}, C.-P., {McConnell}, N.~J., {et~al.} 2019, \apj, 878, 57,
  \dodoi{10.3847/1538-4357/ab1f04}

\bibitem[{{Ene} {et~al.}(2020){Ene}, {Ma}, {Walsh}, {Greene}, {Thomas}, \&
  {Blakeslee}}]{Eneetal2020}
{Ene}, I., {Ma}, C.-P., {Walsh}, J.~L., {et~al.} 2020, \apj, 891, 65,
  \dodoi{10.3847/1538-4357/ab7016}

\bibitem[{Ene {et~al.}(2018)Ene, Ma, Veale, Greene, Thomas, Blakeslee, Foster,
  Walsh, Ito, \& Goulding}]{Eneetal2018}
Ene, I., Ma, C.-P., Veale, M., {et~al.} 2018, \mnras, 479, 2810,
  \dodoi{10.1093/mnras/sty1649}

\bibitem[{Feldmeier-Krause {et~al.}(2017)Feldmeier-Krause, Zhu, Neumayer,
  van~de Ven, de~Zeeuw, \& Sch??del}]{FeldmeierKrause2017}
Feldmeier-Krause, A., Zhu, L., Neumayer, N., {et~al.} 2017, \mnras, 466, 4040,
  \dodoi{10.1093/mnras/stw3377}

\bibitem[{{Foster} {et~al.}(2017){Foster}, {van de Sande}, {D'Eugenio},
  {Cortese}, {McDermid}, {Bland-Hawthorn}, {Brough}, {Bryant}, {Croom},
  {Goodwin}, {Konstantopoulos}, {Lawrence}, {L{\'o}pez-S{\'a}nchez}, {Medling},
  {Owers}, {Richards}, {Scott}, {Taranu}, {Tonini}, \&
  {Zafar}}]{Fosteretal2017}
{Foster}, C., {van de Sande}, J., {D'Eugenio}, F., {et~al.} 2017, \mnras, 472,
  966, \dodoi{10.1093/mnras/stx1869}

\bibitem[{{Franx}(1988)}]{Franx1988}
{Franx}, M. 1988, \mnras, 231, 285, \dodoi{10.1093/mnras/231.2.285}

\bibitem[{{Franx} {et~al.}(1991){Franx}, {Illingworth}, \& {de
  Zeeuw}}]{Franxetal1991}
{Franx}, M., {Illingworth}, G., \& {de Zeeuw}, T. 1991, \apj, 383, 112,
  \dodoi{10.1086/170769}

\bibitem[{{Gebhardt} {et~al.}(2000){Gebhardt}, {Richstone}, {Kormendy},
  {Lauer}, {Ajhar}, {Bender}, {Dressler}, {Faber}, {Grillmair}, {Magorrian}, \&
  {Tremaine}}]{Gebhardtetal2000a}
{Gebhardt}, K., {Richstone}, D., {Kormendy}, J., {et~al.} 2000, \aj, 119, 1157,
  \dodoi{10.1086/301240}

\bibitem[{{Goullaud} {et~al.}(2018){Goullaud}, {Jensen}, {Blakeslee}, {Ma},
  {Greene}, \& {Thomas}}]{Goullaudetal2018}
{Goullaud}, C.~F., {Jensen}, J.~B., {Blakeslee}, J.~P., {et~al.} 2018, \apj,
  856, 11, \dodoi{10.3847/1538-4357/aab1f3}

\bibitem[{{Gravity Collaboration} {et~al.}(2019){Gravity Collaboration},
  {Abuter}, {Amorim}, {Baub{\"o}ck}, {Berger}, {Bonnet}, {Brandner},
  {Cl{\'e}net}, {Coud{\'e} Du Foresto}, {de Zeeuw}, {Dexter}, {Duvert},
  {Eckart}, {Eisenhauer}, {F{\"o}rster Schreiber}, {Garcia}, {Gao}, {Gendron},
  {Genzel}, {Gerhard}, {Gillessen}, {Habibi}, {Haubois}, {Henning}, {Hippler},
  {Horrobin}, {Jim{\'e}nez-Rosales}, {Jocou}, {Kervella}, {Lacour},
  {Lapeyr{\`e}re}, {Le Bouquin}, {L{\'e}na}, {Ott}, {Paumard}, {Perraut},
  {Perrin}, {Pfuhl}, {Rabien}, {Rodriguez Coira}, {Rousset}, {Scheithauer},
  {Sternberg}, {Straub}, {Straubmeier}, {Sturm}, {Tacconi}, {Vincent}, {von
  Fellenberg}, {Waisberg}, {Widmann}, {Wieprecht}, {Wiezorrek}, {Woillez}, \&
  {Yazici}}]{gravitycollaboration2019}
{Gravity Collaboration}, {Abuter}, R., {Amorim}, A., {et~al.} 2019, \aap, 625,
  L10, \dodoi{10.1051/0004-6361/201935656}

\bibitem[{{Jensen} {et~al.}(2021){Jensen}, {Blakeslee}, {Ma}, {Milne}, {Brown},
  {Cantiello}, {Garnavich}, {Greene}, {Lucey}, {Phan}, {Tully}, \&
  {Wood}}]{Jensenetal2021}
{Jensen}, J.~B., {Blakeslee}, J.~P., {Ma}, C.-P., {et~al.} 2021, arXiv
  e-prints, arXiv:2105.08299.
\newblock \doarXiv{2105.08299}

\bibitem[{{Jin} {et~al.}(2020){Jin}, {Zhu}, {Long}, {Mao}, {Wang}, \& {van de
  Ven}}]{Jinetal2020}
{Jin}, Y., {Zhu}, L., {Long}, R.~J., {et~al.} 2020, \mnras, 491, 1690,
  \dodoi{10.1093/mnras/stz3072}

\bibitem[{{Jin} {et~al.}(2019){Jin}, {Zhu}, {Long}, {Mao}, {Xu}, {Li}, \& {van
  de Ven}}]{Jinetal2019}
---. 2019, \mnras, 486, 4753, \dodoi{10.1093/mnras/stz1170}

\bibitem[{{Joseph} {et~al.}(2001){Joseph}, {Merritt}, {Olling}, {Valluri},
  {Bender}, {Bower}, {Danks}, {Gull}, {Hutchings}, {Kaiser}, {Maran},
  {Weistrop}, {Woodgate}, {Malumuth}, {Nelson}, {Plait}, \&
  {Lindler}}]{Josephetal2001}
{Joseph}, C.~L., {Merritt}, D., {Olling}, R., {et~al.} 2001, \apj, 550, 668,
  \dodoi{10.1086/319781}

\bibitem[{{Krajnovi{\'c}} {et~al.}(2018){Krajnovi{\'c}}, {Cappellari},
  {McDermid}, {Thater}, {Nyland}, {de Zeeuw}, {Falc{\'o}n-Barroso}, {Khochfar},
  {Kuntschner}, {Sarzi}, \& {Young}}]{Krajnovicetal2018}
{Krajnovi{\'c}}, D., {Cappellari}, M., {McDermid}, R.~M., {et~al.} 2018,
  \mnras, 477, 3030, \dodoi{10.1093/mnras/sty778}

\bibitem[{{Leung} {et~al.}(2018){Leung}, {Leaman}, {van de Ven}, {Lyubenova},
  {Zhu}, {Bolatto}, {Falc{\'o}n-Barroso}, {Blitz}, {Dannerbauer}, {Fisher},
  {Levy}, {Sanchez}, {Utomo}, {Vogel}, {Wong}, \& {Ziegler}}]{Leungetal2018}
{Leung}, G. Y.~C., {Leaman}, R., {van de Ven}, G., {et~al.} 2018, \mnras, 477,
  254, \dodoi{10.1093/mnras/sty288}

\bibitem[{Liepold {et~al.}(2020)Liepold, Quenneville, Ma, Walsh, McConnell,
  Greene, \& Blakeslee}]{Liepoldetal2020}
Liepold, C.~M., Quenneville, M.~E., Ma, C.-P., {et~al.} 2020, \apj, 891, 4,
  \dodoi{10.3847/1538-4357/ab6f71}

\bibitem[{{Lipka} \& {Thomas}(2021)}]{LipkaThomas2021}
{Lipka}, M., \& {Thomas}, J. 2021, \mnras, \dodoi{10.1093/mnras/stab1092}

\bibitem[{{Ma} {et~al.}(2014){Ma}, {Greene}, {McConnell}, {Janish},
  {Blakeslee}, {Thomas}, \& {Murphy}}]{Maetal2014}
{Ma}, C.-P., {Greene}, J.~E., {McConnell}, N., {et~al.} 2014, \apj, 795, 158,
  \dodoi{10.1088/0004-637X/795/2/158}

\bibitem[{{McConnell} \& {Ma}(2013)}]{mcconnellma2013}
{McConnell}, N.~J., \& {Ma}, C.-P. 2013, \apj, 764, 184,
  \dodoi{10.1088/0004-637X/764/2/184}

\bibitem[{McKay {et~al.}(1979)McKay, Beckman, \& Conover}]{McKayetal1979}
McKay, M.~D., Beckman, R.~J., \& Conover, W.~J. 1979, Technometrics, 21, 239.
\newblock \url{http://www.jstor.org/stable/1268522}

\bibitem[{Moza(2020)}]{LHSMDU}
Moza, S. 2020, {sahilm89/lhsmdu: Latin Hypercube Sampling with
  Multi-Dimensional Uniformity (LHSMDU): Speed Boost minor compatibility
  fixes}, 1.1.1,  Zenodo, \dodoi{10.5281/zenodo.3929531}

\bibitem[{Pedregosa {et~al.}(2011)Pedregosa, Varoquaux, Gramfort, Michel,
  Thirion, Grisel, Blondel, Prettenhofer, Weiss, Dubourg, Vanderplas, Passos,
  Cournapeau, Brucher, Perrot, \& Duchesnay}]{scikit-learn}
Pedregosa, F., Varoquaux, G., Gramfort, A., {et~al.} 2011, Journal of Machine
  Learning Research, 12, 2825

\bibitem[{{Poci} {et~al.}(2019){Poci}, {McDermid}, {Zhu}, \& {van de
  Ven}}]{Pocietal2019}
{Poci}, A., {McDermid}, R.~M., {Zhu}, L., \& {van de Ven}, G. 2019, \mnras,
  487, 3776, \dodoi{10.1093/mnras/stz1154}

\bibitem[{{Quenneville} {et~al.}(2021){Quenneville}, {Liepold}, \&
  {Ma}}]{Quennevilleetal2021}
{Quenneville}, M.~E., {Liepold}, C.~M., \& {Ma}, C.-P. 2021, \apjs, 254, 25,
  \dodoi{10.3847/1538-4365/abe6a0}

\bibitem[{{Schwarzschild}(1993)}]{Schwarzschild1993}
{Schwarzschild}, M. 1993, \apj, 409, 563, \dodoi{10.1086/172687}

\bibitem[{Seth {et~al.}(2014)Seth, van~den Bosch, Mieske, Baumgardt, Brok,
  Strader, Neumayer, Chilingarian, Hilker, McDermid, Spitler, Brodie, Frank, \&
  Walsh}]{Sethetal2014}
Seth, A.~C., van~den Bosch, R., Mieske, S., {et~al.} 2014, \nat, 513, 398,
  \dodoi{10.1038/nature13762}

\bibitem[{{Shapiro} {et~al.}(2006){Shapiro}, {Cappellari}, {de Zeeuw},
  {McDermid}, {Gebhardt}, {van den Bosch}, \& {Statler}}]{Shapiroetal2006}
{Shapiro}, K.~L., {Cappellari}, M., {de Zeeuw}, T., {et~al.} 2006, \mnras, 370,
  559, \dodoi{10.1111/j.1365-2966.2006.10537.x}

\bibitem[{{van den Bosch} \& {de Zeeuw}(2010)}]{vandenBoschdeZeeuw2010}
{van den Bosch}, R.~C.~E., \& {de Zeeuw}, P.~T. 2010, \mnras, 401, 1770,
  \dodoi{10.1111/j.1365-2966.2009.15832.x}

\bibitem[{{van den Bosch} \& {van de Ven}(2009)}]{vandenBoschvandeVen2009}
{van den Bosch}, R. C.~E., \& {van de Ven}, G. 2009, \mnras, 398, 1117,
  \dodoi{10.1111/j.1365-2966.2009.15177.x}

\bibitem[{{van den Bosch} {et~al.}(2008){van den Bosch}, {van de Ven},
  {Verolme}, {Cappellari}, \& {de Zeeuw}}]{vandenBoschetal2008}
{van den Bosch}, R.~C.~E., {van de Ven}, G., {Verolme}, E.~K., {Cappellari},
  M., \& {de Zeeuw}, P.~T. 2008, \mnras, 385, 647,
  \dodoi{10.1111/j.1365-2966.2008.12874.x}

\bibitem[{{van der Marel} {et~al.}(1998){van der Marel}, {Cretton}, {de Zeeuw},
  \& {Rix}}]{vanderMareletal1998}
{van der Marel}, R.~P., {Cretton}, N., {de Zeeuw}, P.~T., \& {Rix}, H.-W. 1998,
  \apj, 493, 613, \dodoi{10.1086/305147}

\bibitem[{{Vasiliev} \& {Valluri}(2020)}]{VasilievValluri2020}
{Vasiliev}, E., \& {Valluri}, M. 2020, \apj, 889, 39,
  \dodoi{10.3847/1538-4357/ab5fe0}

\bibitem[{{Veale} {et~al.}(2017{\natexlab{a}}){Veale}, {Ma}, {Greene},
  {Thomas}, {Blakeslee}, {McConnell}, {Walsh}, \& {Ito}}]{Vealeetal2017b}
{Veale}, M., {Ma}, C.-P., {Greene}, J.~E., {et~al.} 2017{\natexlab{a}}, \mnras,
  471, 1428, \dodoi{10.1093/mnras/stx1639}

\bibitem[{{Veale} {et~al.}(2018){Veale}, {Ma}, {Greene}, {Thomas}, {Blakeslee},
  {Walsh}, \& {Ito}}]{Vealeetal2018}
---. 2018, \mnras, 473, 5446, \dodoi{10.1093/mnras/stx2717}

\bibitem[{{Veale} {et~al.}(2017{\natexlab{b}}){Veale}, {Ma}, {Thomas},
  {Greene}, {McConnell}, {Walsh}, {Ito}, {Blakeslee}, \&
  {Janish}}]{Vealeetal2017a}
{Veale}, M., {Ma}, C.-P., {Thomas}, J., {et~al.} 2017{\natexlab{b}}, \mnras,
  464, 356, \dodoi{10.1093/mnras/stw2330}

\bibitem[{{Verolme} {et~al.}(2002){Verolme}, {Cappellari}, {Copin}, {van der
  Marel}, {Bacon}, {Bureau}, {Davies}, {Miller}, \& {de
  Zeeuw}}]{Verolmeetal2002}
{Verolme}, E.~K., {Cappellari}, M., {Copin}, Y., {et~al.} 2002, \mnras, 335,
  517, \dodoi{10.1046/j.1365-8711.2002.05664.x}

\bibitem[{{Walsh} {et~al.}(2012){Walsh}, {van den Bosch}, {Barth}, \&
  {Sarzi}}]{Walshetal2012}
{Walsh}, J.~L., {van den Bosch}, R.~C.~E., {Barth}, A.~J., \& {Sarzi}, M. 2012,
  \apj, 753, 79, \dodoi{10.1088/0004-637X/753/1/79}

\bibitem[{{Walsh} {et~al.}(2015){Walsh}, {van den Bosch}, {Gebhardt},
  {Yildirim}, {G{\"u}ltekin}, {Husemann}, \& {Richstone}}]{Walshetal2015}
{Walsh}, J.~L., {van den Bosch}, R.~C.~E., {Gebhardt}, K., {et~al.} 2015, \apj,
  808, 183, \dodoi{10.1088/0004-637X/808/2/183}

\bibitem[{{Walsh} {et~al.}(2017){Walsh}, {van den Bosch}, {Gebhardt},
  {Y{\i}ld{\i}r{\i}m}, {G{\"u}ltekin}, {Husemann}, \&
  {Richstone}}]{Walshetal2017}
---. 2017, \apj, 835, 208, \dodoi{10.3847/1538-4357/835/2/208}

\bibitem[{{Walsh} {et~al.}(2016){Walsh}, {van den Bosch}, {Gebhardt},
  {Y{\i}ld{\i}r{\i}m}, {Richstone}, {G{\"u}ltekin}, \&
  {Husemann}}]{Walshetal2016}
---. 2016, \apj, 817, 2, \dodoi{10.3847/0004-637X/817/1/2}

\bibitem[{{Weijmans} {et~al.}(2014){Weijmans}, {de Zeeuw}, {Emsellem},
  {Krajnovi{\'c}}, {Lablanche}, {Alatalo}, {Blitz}, {Bois}, {Bournaud},
  {Bureau}, {Cappellari}, {Crocker}, {Davies}, {Davis}, {Duc}, {Khochfar},
  {Kuntschner}, {McDermid}, {Morganti}, {Naab}, {Oosterloo}, {Sarzi}, {Scott},
  {Serra}, {Verdoes Kleijn}, \& {Young}}]{Weijmansetal2014}
{Weijmans}, A.-M., {de Zeeuw}, P.~T., {Emsellem}, E., {et~al.} 2014, \mnras,
  444, 3340, \dodoi{10.1093/mnras/stu1603}

\bibitem[{{Yang} {et~al.}(2020){Yang}, {Zhu}, {Weijmans}, {van de Ven},
  {Boardman}, {Morganti}, \& {Oosterloo}}]{Yangetal2020}
{Yang}, M., {Zhu}, L., {Weijmans}, A.-M., {et~al.} 2020, \mnras, 491, 4221,
  \dodoi{10.1093/mnras/stz3293}

\bibitem[{{Zhu} {et~al.}(2018{\natexlab{a}}){Zhu}, {van den Bosch}, {van de
  Ven}, {Lyubenova}, {Falc{\'o}n-Barroso}, {Meidt}, {Martig}, {Shen}, {Li},
  {Yildirim}, {Walcher}, \& {Sanchez}}]{Zhuetal2018a}
{Zhu}, L., {van den Bosch}, R., {van de Ven}, G., {et~al.} 2018{\natexlab{a}},
  \mnras, 473, 3000, \dodoi{10.1093/mnras/stx2409}

\bibitem[{{Zhu} {et~al.}(2018{\natexlab{b}}){Zhu}, {van de Ven}, {van den
  Bosch}, {Rix}, {Lyubenova}, {Falc{\'o}n-Barroso}, {Martig}, {Mao}, {Xu},
  {Jin}, {Obreja}, {Grand }, {Dutton}, {Macci{\`o}}, {G{\'o}mez}, {Walcher},
  {Garc{\'\i}a-Benito}, {Zibetti}, \& {S{\'a}nchez}}]{Zhuetal2018b}
{Zhu}, L., {van de Ven}, G., {van den Bosch}, R., {et~al.} 2018{\natexlab{b}},
  Nature Astronomy, 2, 233, \dodoi{10.1038/s41550-017-0348-1}

\end{thebibliography}
\bibliographystyle{aasjournal}

\end{document}